\documentclass[11pt]{article}

\makeatletter
\renewcommand\section{\@startsection {section}{1}{\z@}%
                                   {-3.5ex \@plus -1ex \@minus -.2ex}%
                                   {2.3ex \@plus.2ex}%
                                   {\normalfont\fontfamily{phv}\fontsize{16}{19}\bfseries}}
\renewcommand\subsection{\@startsection{subsection}{2}{\z@}%
                                     {-3.25ex\@plus -1ex \@minus -.2ex}%
                                     {1.5ex \@plus .2ex}%
                                     {\normalfont\fontfamily{phv}\fontsize{14}{17}\bfseries}}
\renewcommand\subsubsection{\@startsection{subsubsection}{3}{\z@}%
                                    {-3.25ex\@plus -1ex \@minus -.2ex}%
                                     {1.5ex \@plus .2ex}%
                                     {\normalfont\normalsize\fontfamily{phv}\fontsize{14}{17}\selectfont}}
\makeatother

\usepackage{amsmath}
\usepackage{amsfonts}
\usepackage{amssymb}

\usepackage{amsthm}
\usepackage{array}
\usepackage{booktabs}
\usepackage{multirow}
\usepackage{graphicx}
\usepackage{threeparttable}
\usepackage{textcomp}
\usepackage{url}
\usepackage{xcolor}
\usepackage{cite}
\usepackage{caption}
\usepackage{microtype}
\usepackage[hidelinks]{hyperref}
\usepackage[all]{hypcap}

\newcommand{\subtextbf}[1]{\textbf{#1}} 

\newcommand{\sepehr}[1]{}
\newcommand{\babak}[1]{}

\newtheorem{proposition}{Proposition}

\begin{document}
\def\spacingset#1{\renewcommand{\baselinestretch}%
    {#1}\small\normalsize} \spacingset{1.5}

\title{When Networks Substitute for Outcome Surveillance?\\[2pt] \large A Substitution–Complementarity Framework for Behavioral Signals in Predictive Monitoring}

\author{Sepehr~Ilami,
        Qingtao~Cao,
        and~Babak~Heydari*%
\thanks{S. Ilami, Q. Cao, and B. Heydari are with the College of Engineering
and Network Science Institute, Northeastern University, Boston, MA 02115 USA.
E-mail: b.heydari@northeastern.edu}%
\thanks{Manuscript received in June 2026}}

\date{}
\maketitle

\begin{abstract}
Monitoring systems increasingly fuse a dynamic behavioral data stream (human mobility, transaction flows, communication patterns) with outcome-based surveillance, raising a basic value-of-information question: when does the behavioral signal carry predictive information that the outcome history does not already contain? We study this question using epidemic forecasting on mobility networks as a testbed, asking whether a town's network position provides \emph{independent} predictive signal (beyond what local outcome-based surveillance already encodes) with direct consequences for the design of network-aware monitoring systems. We formalize this as a \emph{substitution--complementarity} problem over directed, weighted mobility networks and develop an analytical framework, grounded in a Frisch--Waugh--Lovell variance decomposition, that derives domain-agnostic conditions under which network-topology features retain incremental explanatory power beyond autoregressive outcome histories and conditions under which they do not. We instantiate the framework using town-level COVID-19 forecasting in Massachusetts (April 2020--April 2021), constructing directed mobility networks among over 300 towns from weekly smartphone-derived origin--destination aggregates and extracting betweenness centrality and intra-town flow as the primary holistic and local topology features. We then establish generality with an agent-based model in which agents move, interact, and transmit on synthetic networks with known structure, confirming that the regime boundary arises from a generic interaction between macro-scale epidemic state and network topology rather than dataset-specific artifacts. Prevalence-gated interactions between statewide incidence and network-position features yield large out-of-sample gains when the primary surveillance channel is degraded (Predict-$R^2$: ${\sim}0.60\rightarrow{\sim}0.83$--$0.89$) but only marginal lift when granular local histories are available (${\approx}{+}0.5$ percentage points), with gains concentrating during epidemic waves when endogenous behavioral response shifts network connectivity most rapidly. Framed as a \emph{value-of-information} problem, the substitution gain is the marginal value of the behavioral data stream as a function of primary-channel quality, yielding a transferable, cost-aware design rule for predictive monitoring systems: activate topology-aware behavioral-signal integration when primary surveillance is degraded or the network is in a rapid-change phase; otherwise, rely on autoregressive baselines and retain network features as diagnostic overlays for structural monitoring.
\end{abstract}


\section{Introduction}\label{sec:introduction}
Across domains, monitoring systems increasingly pair a primary, outcome-based surveillance channel with an auxiliary stream of dynamic behavioral data (human mobility, transaction flows, communication patterns) whose network structure encodes how risk propagates, and a recurring design question is when that behavioral signal carries predictive information beyond what the outcome history already provides. Epidemic spreading on mobility networks is a natural testbed: a town that bridges two otherwise weakly connected communities occupies a structurally privileged position, its betweenness centrality concentrates transmission pathways, and changes in its inflow or outflow reverberate through the network in ways that aggregate flow volumes alone do not capture. Understanding when these topological properties provide \emph{operationally useful} predictive signal is a core problem in network epidemiology, one that connects the classical theory of spreading processes on complex networks \cite{Barrat2004,Colizza2007,balcan2009multiscale,Brockmann2013,keeling2005networks,butler2024analysis,hota2021closedloop, pare2017epidemic} to the practical design of data-driven monitoring systems. Cast in the language of analytics, this is a \emph{value-of-information} question \cite{howard1966information} (when does fusing an auxiliary, dynamically evolving data stream (here, human mobility \cite{gonzalez2008understanding}) with outcome-based surveillance improve prediction enough to justify its acquisition and integration cost?); one that sits at the predictive-to-prescriptive analytics interface \cite{bertsimas2020predictive,elmachtoub2022smart} and connects to a growing body of work on modeling and monitoring evolving networks \cite{wu2024modeling}.
 
The theoretical literature has established clearly that network structure conditions epidemic thresholds and shapes propagation pathways: the spectral radius of the contact matrix governs the transition between disease-free and endemic states \cite{pare2017epidemic,you2025competitive}, and hub-and-bridge positions concentrate transmission risk under heterogeneous mixing \cite{Barrat2004,Colizza2007,balcan2009multiscale}. What has received far less attention is the \emph{informational} corollary: under what surveillance conditions does the network's structural state carry forecasting value that \emph{outcome-based histories do not already subsume}? This question is consequential because agents being monitored by the system adapt their behavior (mobility, mixing, travel) in response to the very conditions the system tracks \cite{olguin2008sensible,eagle2009inferring,wang2011human, stopczynski2014measuring,zheng2016big,zhang2018urban,wang2025exploring}. This endogenous behavioral response generates dynamic data streams (mobility traces, origin--destination flows, time-varying contact networks) that are increasingly fused into operational forecasting pipelines supporting staffing, resource pre-positioning, and adaptive policy \cite{latif2020leveraging,adiga2021all,lopez2024challenges, ribeiro2020short}. Unlike static structural covariates or outcome-oriented surveillance records, these behavioral signals capture how the network itself evolves in response to risk and policy, making their value conditional on what the primary surveillance channel already provides.
 
Incorporating dynamic network features is not cost-free. Constructing time-varying mobility networks, computing centralities at operational cadence, and maintaining interaction features adds latency, storage, and maintenance burden. Collection of the underlying behavioral traces entails privacy and governance overhead \cite{stopczynski2014measuring,grantz2020use}. Model complexity rises, potentially reducing interpretability. These trade-offs mean that demonstrating \emph{correlation} between a network-topology feature and an outcome of interest is insufficient for system design; what is needed is a characterization of the \emph{conditions} under which the feature's marginal predictive value justifies its marginal cost, particularly when outcome-based autoregressive baselines may already absorb much of the network's cross-sectional signal.
 
This design question (when does a network-topology feature \emph{substitute} for an inadequate primary surveillance channel versus merely \emph{complement} an already-informative one?) recurs across networked sociotechnical domains. In \emph{supply-chain disruption forecasting}, real-time freight network topology (which ports or hubs act as bridges) becomes a critical substitute for missing inventory visibility when supplier reporting lags, as occurred during COVID-19 supply shocks and semiconductor shortages \cite{zhao2018supply,brintrup2016understanding}. In \emph{financial systemic-risk monitoring}, interbank transaction flows (a form of financial mobility network) reveal stress-propagation pathways that balance-sheet data obscure during crises, precisely because price data become noisy and the \emph{network} retains structural signal \cite{acemoglu2015systemic,glasserman2015contagion, brunetti2019interconnectedness}. In each case, the behavioral network is most informative when the primary channel is degraded, and the value of structural features is modulated by the network topology through which the monitored process propagates.

Constructing a universal theoretical framework for this substitution--complementarity boundary would require strong assumptions about signal structure, noise properties, and decision-loss functions that vary across domains. A more productive approach is to develop a rigorous \emph{empirical template}: select a setting in which (a)~both the primary surveillance channel and the dynamic behavioral signal are well-measured over a sustained period, (b)~natural variation in data availability creates identifiable regime contrasts, (c)~the behavioral signal has a rich enough network structure to distinguish volume from topology, and (d)~sufficient cross-sectional and temporal variation exists to evaluate out-of-sample predictive performance under blocked designs. A setting that satisfies all four criteria enables conclusions that, while grounded in one domain, illuminate the general mechanism and provide a transferable analytical template.

Epidemic forecasting satisfies all four criteria simultaneously. Short-horizon forecasts underpin near-term public-health operations and must be produced under tight latency constraints, so the cost of unnecessary behavioral-signal computation is operationally real \cite{lipsitch2020defining,lopez2024challenges, ribeiro2020short}. During COVID-19, aggregated smartphone mobility traces became one of the most widely adopted dynamic behavioral signals in forecasting pipelines, providing a real-time measurement layer on population mixing and behavioral response to risk and policy \cite{grantz2020use,abouk2020immediate,schlosser2020covid,hale2021global,wang2021short,duan2024modeling,li2024trajectory}. Yet empirical findings diverged on \emph{what aspects} of mobility are predictive and, critically, \emph{when} their inclusion is justified given the surveillance data already in hand, making this domain a natural proving ground for the regime-dependent logic outlined above.

Within this domain, we study town-level forecasting in Massachusetts (April 2020--April 2021), a setting that satisfies all four template criteria. The \emph{primary surveillance channel} is the official town-level weekly case counts, whose availability and timeliness varied across the study period, creating a natural regime contrast between data-limited and data-rich conditions \cite{MA_datasets}. The \emph{dynamic behavioral signal} is weekly, directed, weighted inter-town mobility derived from anonymized smartphone origin--destination aggregates \cite{SafeGraph}, which can be encoded both as movement volume (how much flow occurs) and as network topology (which towns occupy bridge-like or hub positions). The cross-sectional scope (over 300 towns), temporal depth (54 weeks spanning two major waves, inter-wave troughs, and policy shifts), and the availability of statewide macro incidence as a shared contextual variable together provide the variation needed for rigorous regime-dependent evaluation. This setting also permits mechanistic triangulation: we complement the reduced-form empirical analysis with an agent-based model (ABM) operating on controlled mobility-like networks, confirming that the observed patterns reflect a general transmission--information mechanism rather than idiosyncrasies of one dataset.

We study the \emph{target variable} $y_{i,t+1}$, the number of new COVID-19 cases in town~$i$ in the next epidemiological week. We evaluate two operational regimes that reflect the substitution-complementarity logic. In a \emph{data-limited regime}, timely town-level case histories may be missing, delayed, or coarsened, so models must rely on macro-level incidence and mobility-derived signals. In a \emph{data-rich regime}, recent town-level cases are available; here, a standard \emph{autoregressive (AR)} baseline augments predictors with the town's most recent cases $y_{i,t}$ to capture short-term persistence. From weekly anonymized origin--destination aggregates, we construct directed mobility networks and extract (i)~mobility-volume proxies (e.g., within-town flow) and (ii)~network-position features (e.g., betweenness centrality capturing bridge-like connectivity). We compare macro-only baselines, mobility-augmented models, and \emph{prevalence-gated interaction models} that couple statewide incidence with mobility-derived features, reflecting the hypothesis that the relevance of behavioral signals depends on epidemic context.

Our core empirical question is therefore:
\begin{quote}
\emph{Under which data regimes and epidemic conditions do mobility-derived signals (volume and network structure) provide measurable out-of-sample gains for next week's town-level forecasts beyond standard baselines, including AR models when local histories are available?}
\end{quote}

Two results emerge. First, in the data-limited regime, mobility-derived signals provide large gains, with the strongest improvements coming from prevalence-gated interactions: Predict-$R^2$ increases from ${\sim}0.60$ (macro-only) to ${\sim}0.83$--$0.89$ when interactions between statewide incidence and town-level mobility features are added. Second, in the data-rich regime, AR baselines already capture most of the one-week-ahead predictability (Predict-$R^2 \approx 0.90$), and mobility features deliver only marginal lift (approximately $+0.5$ percentage points). Gains concentrate during the wave and rising-phase periods when both prevalence and connectivity change rapidly. An ABM reproduces the same regime-dependent pattern under controlled dynamics, and an analytical decomposition based on the Frisch--Waugh--Lovell theorem explains why the cross-sectional interaction signal is strong in the macro-only regime but attenuates once recent local histories enter the baseline.

Our contributions, framed for the broader design of sociotechnical systems, are threefold: 1) \textbf{Substitution--complementarity characterization.} We provide the first explicit, regime-based quantification of when a dynamic behavioral signal (mobility) substitutes for degraded primary surveillance versus offering only marginal complementarity to an already-informative channel, a question that generalizes beyond epidemiology to any sociotechnical monitoring system that fuses behavioral and outcome data. 2) \textbf{Encoding matters conditionally.} We show that the appropriate encoding of the behavioral signal (volume, network topology, or prevalence-gated interaction) is itself regime-dependent, not universally optimal. Prevalence-gated interactions dominate in the data-limited regime; simpler AR baselines suffice in the data-rich regime. 3) \textbf{Conditional-module system architecture.} We translate the findings into a prescriptive system architecture in which behavioral-signal integration is activated or deactivated based on the quality of the primary surveillance channel and the dynamic state of the monitored process. Explicit triggering conditions follow from the analytical and simulation results: invest in real-time behavioral-signal computation when primary surveillance is degraded, or the monitored process is in a rapid-change phase; otherwise, rely on autoregressive baselines and retain behavioral features as a diagnostic overlay for structural monitoring.

The rest of the paper is as follows. Section~\ref{sec:related} reviews related work spanning epidemic processes on complex networks, mobility-informed forecasting, and the substitution--complementarity of behavioral signals in networked monitoring systems. Section~\ref{sec:framework} develops the analytical framework: the forecasting specifications, the Frisch--Waugh--Lovell variance decomposition, and a constructive derivation of the macro--network interaction term. Section~\ref{sec:empirical} presents the empirical validation on Massachusetts town-level forecasting across regimes, epidemic phases, and forecast horizons. Section~\ref{sec:generalizability} establishes generality with an agent-based model spanning multiple network families, heterogeneity levels, endogenous mobility, and channel-degradation conditions. Section~\ref{sec:discussion} discusses mechanisms, limitations, external validity, and implications for system design.


\section{Related Work}\label{sec:related}
 
The question of when dynamic behavioral network signals improve sociotechnical forecasting connects four bodies of literature: (1) epidemic processes on complex networks, (2) mobility-informed epidemic forecasting, (3) the value of auxiliary data in prediction, and (4) substitution and complementarity across networked domains. Table~\ref{tab:related_work} positions our work relative to the most closely related studies across these streams.
 
\begin{table}[!t]
\caption{Comparison of Related Work. Columns indicate whether a study uses dynamic mobility-network topology features, tests incremental value over an autoregressive (AR) baseline, performs regime-dependent evaluation (data-limited vs.\ data-rich), provides a domain-agnostic analytical mechanism, and validates via an agent-based model (ABM).}
\label{tab:related_work}
\centering
\renewcommand{\arraystretch}{1.2}
\resizebox{\linewidth}{!}{%
\begin{tabular}{@{}llccccc@{}}
\toprule
\textbf{Study} & \textbf{Setting} &
\textbf{\shortstack{Network\\topology}} &
\textbf{\shortstack{Tested vs.\\AR baseline}} &
\textbf{\shortstack{Regime\\contrast}} &
\textbf{\shortstack{Analytical\\mechanism}} &
\textbf{ABM} \\
\midrule
Colizza et al.\ \cite{Colizza2007}
  & Global outbreak, metapopulation & \checkmark & $\times$ & $\times$ & \checkmark & $\times$ \\
Balcan et al.\ \cite{balcan2009multiscale}
  & Global air-travel network & \checkmark & $\times$ & $\times$ & \checkmark & $\times$ \\
Schlosser et al.\ \cite{schlosser2020covid}
  & Mobility + COVID-19 (Germany) & \checkmark & $\times$ & $\times$ & $\times$ & \checkmark \\
Chinazzi et al.\ \cite{chinazzi2020effect}
  & Travel restrictions + COVID-19 & \checkmark & $\times$ & $\times$ & $\times$ & \checkmark \\
Duan et al.\ \cite{duan2024modeling}
  & Spatiotemporal contact model & \checkmark & $\times$ & $\times$ & $\times$ & \checkmark \\
Butler et al.\ \cite{butler2024analysis}
  & Networked SEIRS + population flows & \checkmark & $\times$ & $\times$ & \checkmark & $\times$ \\
Paré et al.\ \cite{pare2017epidemic}
  & Epidemic on time-varying networks & \checkmark & $\times$ & $\times$ & \checkmark & $\times$ \\
Hota et al.\ \cite{hota2021closedloop}
  & Closed-loop SIR inference \& control & \checkmark & $\times$ & $\times$ & \checkmark & $\times$ \\
Bjerre-Nielsen et al.\ \cite{BjerreNielsen2021}
  & Behavioral sensing (education) & $\times$ & \checkmark & $\times$ & $\times$ & $\times$ \\
Wang et al.\ \cite{wang2021short}
  & COVID-19 forecasting (urban) & \checkmark & $\times$ & $\times$ & $\times$ & $\times$ \\
Xu et al.\ \cite{xu2024inferring}
  & Mobility inference (sparse data) & $\times$ & $\times$ & $\times$ & $\times$ & $\times$ \\
\midrule
\textbf{This work}
  & Epidemic forecasting (MA towns) & \checkmark & \checkmark & \checkmark & \checkmark & \checkmark \\
\bottomrule
\end{tabular}}
\end{table}
 
\subsection{Epidemic Processes on Complex Networks}
A well-established body of theory characterises how network topology governs epidemic thresholds and spreading dynamics~\cite{Barrat2004,Colizza2007,balcan2009multiscale, Brockmann2013,keeling2005networks,newman2002spread, PastorSatorras2001,meyers2007contact,kiss2017mathematics}. Colizza et al.\ \cite{Colizza2007} showed that heterogeneous metapopulation networks couple local epidemic trajectories through mobility flows, with hub nodes acting as primary sources of long-range spread. Balcan et al.\ \cite{balcan2009multiscale} demonstrated that the topology of global mobility networks modulates both the speed and geographic pattern of pandemic emergence. Brockmann and Helbing~\cite{Brockmann2013} revealed that effective-distance geometry (derived from directed flow networks) better predicts arrival times than physical distance, underscoring the primacy of network structure over geography.
 
Several recent contributions have extended this tradition to the prediction and control of epidemic processes on time-varying and flow-coupled networks. Paré et al.\ \cite{pare2017epidemic} established stability conditions for SIS epidemic processes on continuous-time time-varying networks. Hota et al.\ \cite{hota2021closedloop} developed a closed-loop framework for simultaneous inference and control of SIR epidemics on networks, exploiting graph structure to separate estimation uncertainty from control effort. Butler et al.\ \cite{butler2024analysis} formulated a networked SEIRS model that treats inter-subpopulation population flows as the propagation mechanism for infection, deriving stability conditions and applying the framework to COVID-19 travel and infection data from counties in Minnesota. You et al.\ \cite{you2025competitive} analyzed optimal control of competitive epidemic spreading processes with memory on heterogeneous networks, showing that network heterogeneity creates non-trivial tradeoffs between containment and resource allocation.
 
Our work complements this line of research by asking not how network structure governs epidemic dynamics under \emph{known} contact matrices, but rather when network-derived features provide \emph{incremental predictive value} for empirical forecasting beyond what autoregressive outcome histories already capture, a question not addressed by existing theoretical or control-theoretic treatments.

\subsection{Mobility-Informed Epidemic Forecasting}
Mobility data in epidemic modeling can be encoded as \emph{intensity} (origin--destination counts, dwell times, aggregated indices \cite{pinheiro2021using,duan2024modeling}) or as \emph{structure}, time-varying network features such as centralities and bridge positions that modulate diffusion pathways \cite{keeling2005networks,newman2002spread, PastorSatorras2001,meyers2007contact,kiss2017mathematics}. Weighted mobility networks condition spreading dynamics through both volume and topology \cite{Barrat2004,Colizza2007,balcan2009multiscale,Brockmann2013}, and macro-level social connectivity modulates risk perception, policy responses, and pandemic trajectories \cite{chinazzi2020effect,milani2021covid}. During COVID-19, smartphone-derived origin--destination aggregates became the dominant behavioral proxy for inter-community mixing \cite{grantz2020use,schlosser2020covid,hale2021global}.
 
Recent work has developed spatiotemporal contact models for forecasting \cite{duan2024modeling,li2024trajectory,wang2021short, lu2026human} and explored mobility inference under sparse observations \cite{xu2024inferring}. However, the dominant evaluation protocol compares mobility-augmented models to static baselines or minimal null models, rather than testing incremental gains over the strongest available outcome-based benchmark, the autoregressive model that uses the target jurisdiction's own recent case history. As a consequence, reported improvements conflate genuine network-structural signal with the predictability already embedded in recent outcomes. Even some recent studies that exploit urban mobility-network topology for epidemic containment \cite{cheng2025poi} optimize network-level interventions without evaluating whether topology features add forecasting value conditional on granular local case histories, a gap shared across the field. Our work fills this gap by adopting the autoregressive model as the data-rich baseline and measuring the precise margin by which network-topology features extend or fail to extend predictability.

\subsection{Value of Auxiliary Behavioral Data in Prediction} 
Surprisingly few studies address the incremental value of behavioral streams conditional on primary surveillance data. Bjerre-Nielsen et al.\ \cite{BjerreNielsen2021} provide a rare exception: in an educational setting, they show that task-specific outcome histories outperform broad behavioral traces, concluding that the value of behavioral sensing depends on the availability of outcome data. Their finding foreshadows the regime-dependent pattern we document in a network-epidemic setting. More broadly, the question of whether auxiliary behavioral signals \emph{substitute} for degraded primary surveillance or merely \emph{complement} strong surveillance arises (though rarely framed explicitly) across networked sociotechnical domains. In supply-chain management, network topology modulates how disruptions propagate, and flow data becomes most valuable when supplier reporting lags \cite{zhao2018supply,brintrup2016understanding}. In financial systemic-risk monitoring, interbank transaction-network topology reveals stress-propagation pathways that balance-sheet data obscure during crises \cite{acemoglu2015systemic, glasserman2015contagion,brunetti2019interconnectedness}. Despite this recurring pattern, no study has empirically quantified the substitution--complementarity boundary with explicit regime contrasts, an out-of-sample evaluation protocol, and a
domain-agnostic analytical mechanism. Our work provides all three.
 
\subsection{Summary of Gaps Addressed}
Table~\ref{tab:related_work} makes the gap precise. The network-epidemic literature establishes topology's role in spreading dynamics, but treats contact matrices as known rather than asking when topology-derived features carry forecasting value beyond observed outcomes. The mobility-forecasting literature reports improvements over static baselines. Still, it does not isolate the network-structural signal from the outcome-history signal, and none of these works derive a domain-agnostic analytical mechanism for the regime boundary. We address all three gaps within a single integrated framework, with mechanistic validation using an agent-based model.

\section{Analytical Framework}
\label{sec:framework}

We formalize the substitution--complementarity problem by asking when a dynamic
behavioral network signal provides incremental predictive value beyond the
information already contained in outcome histories. The central issue is not
whether mobility networks shape epidemic dynamics, a point well established in
network epidemiology, but whether mobility-derived features contain forecasting
information that has not already been absorbed by recent local case counts.

We proceed in three steps. First, we define the forecasting specifications and
the two operational regimes. Second, we use a partialling-out argument based on
the Frisch--Waugh--Lovell theorem to clarify why macro--network interactions can
have large value when local histories are unavailable but only marginal value
when autoregressive histories are present. Third, we derive the macro--network
interaction term from a minimal networked transmission process. This derivation
shows that prevalence-gated network exposure is not an ad hoc feature: it is the
leading aggregate component of transmission on a mobility network.

\subsection{Forecasting Specifications and Operational Regimes}

Let $y_{i,t}$ denote new cases in town $i$ during epidemiological week $t$, and
let $MA_t$ denote statewide incidence during the same week. Let $X_{i,t}$ collect
mobility-derived features for town $i$ at time $t$, such as local flow volume,
inflow strength, or network-position measures. All specifications include town
fixed effects $\alpha_i$.

The baseline data-limited specification uses only macro-level incidence:
\begin{equation}
    y_{i,t+1} = \alpha_i + \beta MA_t + \varepsilon_{i,t+1}.
    \label{eq:basic}
\end{equation}

We then augment this baseline with mobility-derived features and with
prevalence-gated interactions:
\begin{equation}
    y_{i,t+1}
    =
    \alpha_i
    + \beta MA_t
    + \gamma^\top X_{i,t}
    + \delta^\top (MA_t X_{i,t})
    + \varepsilon_{i,t+1}.
    \label{eq:non_ar}
\end{equation}

The interaction term $MA_t X_{i,t}$ represents the idea that the predictive
value of a town's network position depends on the macro-scale epidemic context.
A bridge-like town is not equally informative when statewide prevalence is low
and when statewide prevalence is high.

When timely local case histories are available, the data-rich specification adds
an autoregressive term:
\begin{equation}
    y_{i,t+1}
    =
    \alpha_i
    + \phi y_{i,t}
    + \beta MA_t
    + \gamma^\top X_{i,t}
    + \delta^\top (MA_t X_{i,t})
    + \varepsilon_{i,t+1}.
    \label{eq:ar}
\end{equation}

Equations~\eqref{eq:non_ar} and~\eqref{eq:ar} define the two operational regimes.
In the data-limited regime, town-level recent cases are missing, delayed, or too
coarsened for reliable use, so behavioral network signals may substitute for
degraded local surveillance. In the data-rich regime, recent town-level outcomes
are observed, and mobility features can only complement an already-informative
autoregressive baseline.

\subsection{Incremental Value as Partial Explainability}
\label{sec:fwl}

The incremental value of the interaction term can be understood through a
partialling-out argument. Consider a single network exposure feature $Z_{i,t}$
and the interaction
\[
    W_{i,t} = MA_t Z_{i,t}.
\]
After removing the baseline regressors, the contribution of $W_{i,t}$ is governed
by the covariance between the residualized outcome and the residualized
interaction. Equivalently, by the Frisch--Waugh--Lovell theorem, the partial
$R^2$ of $W_{i,t}$ is the explanatory power obtained by regressing the residual
outcome on the residual component of $W_{i,t}$ that is orthogonal to the
baseline. This residual-on-residual logic is the classical antecedent of modern
partialling-out (double/debiased machine-learning) estimators
\cite{chernozhukov2018double}, situating the decomposition within contemporary
predictive analytics.

This observation clarifies the substitution--complementarity boundary. In the
data-limited regime, the baseline contains town fixed effects and macro
incidence, but not recent town outcomes. The residual component of
$MA_t Z_{i,t}$ therefore retains substantial cross-town variation in network
exposure, especially in weeks when statewide prevalence is high. In this regime,
the interaction can recover cross-sectional information about where aggregate
epidemic pressure is likely to be expressed through the mobility network.

In the data-rich regime, however, the baseline also contains $y_{i,t}$. Recent
local incidence is itself an outcome of prior transmission, mobility, and local
susceptibility. It therefore encodes much of the near-term structural signal
that the network feature is trying to reconstruct. After partialling out
$y_{i,t}$, the remaining component of $MA_t Z_{i,t}$ is smaller, and its
incremental explanatory value is correspondingly attenuated.

This logic is general. It applies whenever a common contextual variable, such as
macro prevalence, interacts with a cross-sectional behavioral feature, such as
network position, while competing with an autoregressive outcome history that
already embeds the recent consequences of that behavior. Thus, the same
mathematical structure can describe other networked monitoring settings: for
example, freight-network topology when inventory reports are delayed, or
transaction-network topology when balance-sheet data are stale.

A useful implication is that, with town fixed effects included, the interaction
is not identified primarily by permanent differences between towns. Instead, its
estimable contribution comes from temporal deviations in a town's mobility
position, or local mixing intensity, interacted with the statewide epidemic
context. The empirical question is therefore whether changes in network exposure
contain predictive information beyond persistent town characteristics and, in
the data-rich regime, beyond the town's own recent case history.

\subsection{Deriving the Macro--Network Interaction}

We now show why a macro--network interaction arises naturally from transmission
on a mobility network. Let $M_t \in \mathbb{R}_{\geq 0}^{n \times n}$ be the
directed, weighted mobility matrix in week $t$, with entry $w_{ij,t}$ denoting
flow from town $i$ to town $j$. Self-loops $w_{ii,t}$ capture within-town
movement or local mixing. Let
\[
    y_t = (y_{1,t},\ldots,y_{n,t})^\top
\]
collect town-level cases.

A minimal linearized transmission step on this network can be written as
\begin{equation}
    y_{t+1} = \beta M_t^\top y_t + \varepsilon_{t+1},
    \label{eq:linear_transmission}
\end{equation}
where $\beta>0$ is an effective transmissibility parameter. This expression is
the leading-order approximation of a networked epidemic process: next-week cases
in a town depend on contact-weighted exposure to infectious individuals in
connected towns.

Decompose the current case vector into its aggregate and mean-zero components:
\[
    y_t = \bar y_t \mathbf{1} + \tilde y_t,
    \qquad
    \bar y_t = \frac{1}{n}\sum_i y_{i,t},
    \qquad
    \mathbf{1}^\top \tilde y_t = 0.
\]
Substituting into~\eqref{eq:linear_transmission} gives
\begin{equation}
    y_{t+1}
    =
    \beta \bar y_t M_t^\top \mathbf{1}
    +
    \beta M_t^\top \tilde y_t
    +
    \varepsilon_{t+1}.
    \label{eq:operator_decomposition}
\end{equation}

Let
\[
    r_t = M_t^\top \mathbf{1}
\]
be the vector of town-level inflow strength. Since $n\bar y_t$ is proportional
to aggregate incidence, and hence to $MA_t$, the first term in
\eqref{eq:operator_decomposition} can be written as
\begin{equation}
    \beta \bar y_t r_t
    =
    \frac{\beta}{n} MA_t r_t.
    \label{eq:macro_network_exact}
\end{equation}

Equation~\eqref{eq:macro_network_exact} is the key theoretical object. It shows
that a macro-incidence-by-network-exposure term emerges directly from applying
the mobility operator to the aggregate component of the epidemic state. In the
one-step linearized process, the exact structural multiplier is inflow strength.
Thus, the interaction is not simply an empirical convenience; it is the leading
aggregate component of network-mediated transmission.

\begin{proposition}[Macro--network interaction as the aggregate transmission component]
\label{prop:decomp}
Under the linearized transmission step~\eqref{eq:linear_transmission}, the
one-step predictor admits the exact decomposition
\[
  y_{t+1}
  \;=\;
  \frac{\beta}{n}\,MA_t\,r_t
  \;+\;
  \beta\,M_t^\top \tilde y_t
  \;+\;
  \varepsilon_{t+1},
  \qquad r_t = M_t^\top \mathbf{1},
\]
in which the first term is a rank-one macro--network interaction whose structural
multiplier is town inflow strength $r_t$. With town fixed effects, this term is
the projection of the transmission operator onto its aggregate, prevalence-scaled
subspace, and it is precisely the component recoverable without observing the
full vector of recent town-level outcomes.
\end{proposition}

Proposition~\ref{prop:decomp} is the analytical foundation of the
substitution--complementarity result, and it is the load-bearing object of the
paper. The feature that the data-limited models exploit is not an arbitrary
modeling choice but the leading aggregate component of network-mediated
transmission, and it is exactly the part of the signal that a town's own recent
case history later renders largely redundant, which is what produces the collapse
of incremental value in the data-rich regime.

The second term in~\eqref{eq:operator_decomposition},
\[
    \beta M_t^\top \tilde y_t,
\]
captures local cross-sectional coupling: how deviations of town-level incidence
from the statewide average are redistributed through the network. This term is
difficult to recover in the data-limited regime because it requires observing
the full vector of recent town-level outcomes. In that setting, the
macro--network interaction in~\eqref{eq:macro_network_exact} represents the
recoverable component of network-mediated exposure. In the data-rich regime,
however, $y_{i,t}$ is observed and already contains recent information about
local deviations, transmission, and network exposure. This is why the
autoregressive term can substantially reduce the residual value of
mobility-derived predictors.

\subsection{From Inflow Strength to Operational Network Features}

The derivation above identifies inflow strength as the exact one-step structural
multiplier. Over longer effective horizons, however, transmission depends on
multi-step movement through the network. Iterating the transmission operator
introduces powers of $M_t^\top$, so the relevant exposure is no longer only
one-step inflow but a walk- or path-based function of network position.

These operator powers motivate the class of \emph{walk- and spectral-exposure}
measures, eigenvector centrality, Katz-like exposure, and PageRank-like
exposure, which are the natural limits of iterated propagation. They do
\emph{not} single out flow betweenness, which is a shortest-path object rather
than a walk or spectral limit. Betweenness instead enters our analysis on
empirical, not derivational, grounds: it is selected as an interpretable,
bridge-mediated representative of network position through an ablation over a
broad candidate set (and is corroborated by the heterogeneity result of
Section~\ref{sec:generalizability}). The theoretical derivation requires only
\emph{some} structural multiplier that captures how aggregate prevalence is
distributed through the mobility network; it does not require betweenness to be
the unique or limiting statistic implied by the model. Betweenness and intra-town
flow are then chosen as empirically strong and interpretable representatives of
this broader class of mobility-derived exposure measures.

This distinction is important. The theory derives the need for a
prevalence-gated network-exposure term. The empirical feature-selection step
then determines which observable mobility summaries best operationalize that
exposure in the Massachusetts town network.

\subsection{Implications for Multi-Week Forecasting}

The same framework also clarifies when mobility-derived predictors should retain
value beyond one-week-ahead forecasts. Suppose mobility evolves according to
\begin{equation}
    M_{t+1} = \Phi(y_t,M_t),
    \label{eq:endogenous_mobility}
\end{equation}
where $\Phi$ represents behavioral response to prevalence, policy, and recent
mobility. Future mobility matrices are then not arbitrary exogenous objects:
they are constrained by the epidemic--mobility trajectory already observed at
time $t$.

If the mobility operator changes slowly over the forecast window, then using
$M_t$ as a proxy for near-future operators introduces limited bias. Formally, if
\[
    \|M_{t+s}-M_t\| \leq \delta
    \qquad
    \text{for } s=1,\ldots,k-1,
\]
and if the operator norms along the path are bounded by $B$, then a telescoping
product bound implies that the difference between the true $k$-step linear
operator and the frozen-operator approximation is bounded on the order of
\begin{equation}
    \beta^k k B^{k-1}\|y_t\|\delta.
    \label{eq:drift_bound}
\end{equation}

\begin{proposition}[Operator-drift bound]
\label{prop:multistep}
Assume the linearized step~\eqref{eq:linear_transmission} with mobility evolving
endogenously as in~\eqref{eq:endogenous_mobility}, per-step operator drift
$\|M_{t+s}-M_t\|\le\delta$ for $s=1,\dots,k-1$, and operator norms bounded by $B$
along the path. Then the frozen-operator $k$-step predictor
$\beta^{k}(M_t^\top)^{k} y_t$ differs from the true $k$-step linear map by at most
$\beta^{k} k\, B^{k-1}\|y_t\|\,\delta$. Multi-step behavior is therefore governed
by the operator drift $\delta$, which the endogenous coupling
\eqref{eq:endogenous_mobility} constrains.
\end{proposition}

Thus, multi-week forecastability depends on operator drift. During stable
periods, the current mobility network can remain informative several weeks
ahead. During rapid-change periods, such as epidemic waves or policy shifts,
operator drift can increase, making frozen-network multi-step forecasts less
stable.

This implication is distinct from the one-week substitution result. Rapid-change
periods may increase the incremental value of current mobility features relative
to an autoregressive baseline because recent cases no longer fully summarize the
current network state. At the same time, rapid operator drift can reduce the
accuracy of multi-week forecasts that assume the current mobility network will
remain stable. The framework therefore predicts two related but separate
patterns: mobility interactions should be most valuable for detecting current
network-mediated risk when the system is changing, while multi-week persistence
of that advantage should be strongest when mobility dynamics are relatively
stable.

We stress that Proposition~\ref{prop:multistep} is a statement about operator
\emph{drift} (the mechanism) rather than a directly testable forecast-bias
prediction. The bound concerns the linearized operator; in a nonlinear epidemic
the realized multi-step error is dominated by linearization error that masks the
drift term. Accordingly, we test the proposition by measuring operator drift
directly under endogenous versus exogenous mobility
(Section~\ref{sec:generalizability}), and we treat the forecast-bias corollary as
illustrative of the mechanism rather than as a quantitative prediction.

\subsection{Summary of Theoretical Predictions}

The analytical framework yields four testable predictions.

First, in the data-limited regime, macro--network interactions should provide
large gains over macro-only baselines because they recover the aggregate
component of network-mediated exposure (Proposition~\ref{prop:decomp}).

Second, in the data-rich regime, the marginal value of network features should
be much smaller because recent town-level cases already encode much of the
near-term transmission and mobility signal.

Third, the value of network features should increase with topological
heterogeneity. If towns occupy similar network positions, the structural
multiplier has little cross-sectional variation; if some towns act as bridges,
hubs, or strong local-mixing centers, network exposure becomes more informative.

Fourth, the usefulness of mobility-derived signals should be phase-dependent.
They should matter most when macro prevalence is high or when the mobility
network is changing in ways not yet fully reflected in local case histories.
Conversely, when local histories are reliable and the system is stable,
autoregressive baselines should dominate, and mobility features should be used
primarily as diagnostic overlays rather than as always-on forecasting modules.

\section{Empirical Validation: Town-Level COVID-19 Forecasting in Massachusetts}\label{sec:empirical}

\textbf{Data and Study Period} We instantiate the substitution–complementarity framework in the following empirical setting, selected to satisfy the template criteria outlined in Section~\ref{sec:introduction}. The pipeline we describe, from raw behavioral traces through network construction, feature extraction, regime detection, and conditional model selection, constitutes a modular monitoring system whose architecture is informed by the analytical and simulation results that follow. To do that, we analyze Massachusetts towns from January 1, 2020, through April 16, 2021. Weekly COVID-19 case counts come from official state and town reports \cite{MA_datasets}. Weekly inter-town mobility is derived from anonymized, aggregated smartphone traces \cite{SafeGraph} from April 13, 2020, through April 16, 2021. Mobility data are pre-processed into origin--destination counts between towns and aggregated by epidemiological week. All mobility data are de-identified and reported at the town level. Analyses comply with relevant data-use agreements and institutional policies.

\textbf{Mobility Traces as a Measurement Signal: Aggregation, Bias, and Normalization} As discussed in Section~I, the value of any dynamic behavioral signal
depends on how faithfully it measures the underlying process of
interest. Smartphone-derived mobility traces provide a behavioral proxy for inter-place mixing but are not a direct measure of contacts. In this study, mobility is used as a \emph{measurement signal} that may correlate with transmission opportunities through repeated co-location, travel-mediated exposure, and changes in mixing induced by policy and risk perception \cite{grantz2020use,schlosser2020covid}. Because raw device-level traces are not observed, we work with aggregated and de-identified origin--destination (OD) counts reported at the town-week level \cite{SafeGraph}. This aggregation improves privacy but introduces modeling considerations: representativeness of the device panel, temporal stability of sampling, and sensitivity to normalization choices.

We address these concerns as follows. All covariates enter forecasting
specifications in standardized form, and features are designed to
capture \emph{relative} mixing and network position rather than
absolute flow magnitudes. Within-town flow (self-loops) serves as a
local mixing proxy; inter-town flows capture coupling between
jurisdictions. In the Supplementary Material, we verify that the main
regime-dependent results are qualitatively stable under alternative
scaling choices, including per-capita normalization and row-normalized
transition shares. These design choices reflect a deployment-oriented
priority: models should be robust to heterogeneous and time-varying
data quality across locations. Measurement imperfections also bear
directly on the substitution--complementarity question: if a noisy
behavioral signal still yields large forecasting gains in the
data-limited regime, this strengthens the case that the underlying
network-position information is genuinely predictive rather than an
artifact of precise measurement.

\begin{figure}[!t]
\centering
\includegraphics[width=0.92\textwidth]{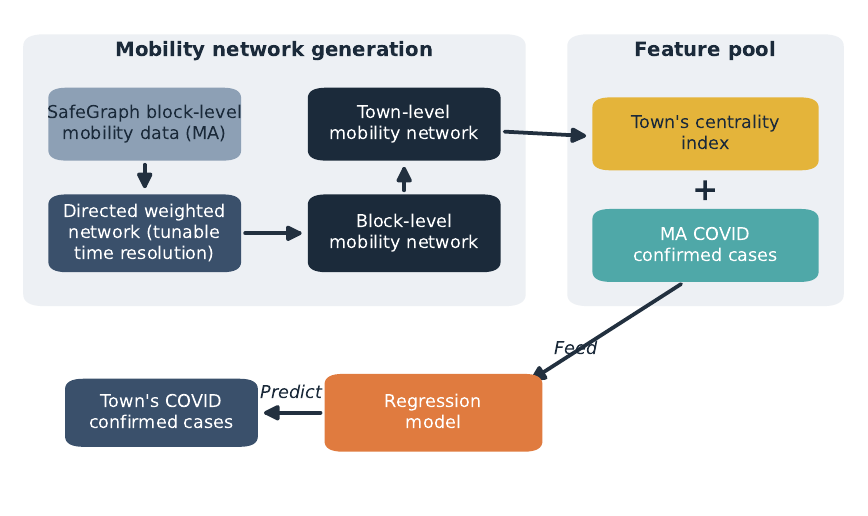}
\caption{Framework overview. Weekly directed, weighted town mobility networks are constructed from SafeGraph origin--destination traces; network-position features (betweenness, intra-town flow) are pooled with statewide incidence $MA_t$ and fed to conditional forecasting models (non-AR vs.\ AR regimes, with prevalence-gated interactions) that predict next-week town-level cases.}
\label{fig:overview}
\end{figure}

\textbf{Constructing the Weekly Mobility Network} Our analytical framework is visually summarized in Fig.~\ref{fig:overview}. For each week $t$, we construct a directed, weighted network $G_t=(V, E_t)$ with nodes $V$ as towns and edge weights $w_{ij,t}$ proportional to the number of device transitions from town $i$ to town $j$. Self-loops $w_{ii,t}$ capture within-town mobility. We compute dynamic topology measures for each $t$, including betweenness centrality (standard flow betweenness on the directed, weighted graph); local flow metrics include intra-town flow $w_{ii,t}$. We selected betweenness (\textit{Bet}) and intra-town flow (\textit{Intraweightt}) as the primary \emph{holistic} and \emph{local} mobility features based on an ablation-style comparison across a broader candidate set (11 holistic centralities and 8 local mobility measures). In these comparisons, \textit{Bet} provided the largest out-of-sample improvement among holistic measures, and \textit{Intraweightt} was the strongest local mobility proxy (see Supplementary Material, Tables S4--S5). This selection step reduces sensitivity to metric choice and anchors the main analysis on the most predictive and interpretable representatives of network position (bridge-like connectivity) and within-town mixing.

Out-of-sample performance is evaluated under a blocked time-wise protocol; details and leakage controls are described in the Supplementary Material.

\subtextbf{Evaluation Protocol and Leakage Control}
We evaluate out-of-sample performance using blocked, time-wise cross-validation to avoid temporal leakage. Each fold holds out a contiguous block of weeks as the test set, and all model fitting (including feature standardization) is performed using only training weeks. Mobility-derived features at week $t$ are constructed solely from mobility observations within week $t$ (no future information), and targets are defined as $y_{i,t+h}$ for horizons $h\in\{1,2,3,4\}$. The primary metric is Predict-$R^2 = 1 - \sum (y-\hat{y})^2 / \sum (y-\bar{y})^2$ computed on held-out observations, where $\bar{y}$ is the mean of the held-out outcomes within the fold. We report Predict-$R^2$ alongside in-sample $R^2$ for reference and emphasize out-of-sample comparisons between model families.

\textbf{Epidemic Phase Definitions and Horizon Analysis} We define \emph{First} and \emph{Second} waves using the standard statewide incidence landmarks and policy chronology, and label \emph{Two Waves} as their union; the \emph{Smooth} period excludes them. Forecast horizons of 1--4 weeks use identical feature sets but adjust the target to $y_{i,t+h}$ for $h\in\{1,2,3,4\}$.

We now evaluate the substitution–complementarity boundary empirically. The analysis proceeds in two regimes (without and with town-level recent cases), using the model families and evaluation protocol described in Section~\ref{sec:framework}. We report Predict-R² and its lift ($\Delta$) over the regime-appropriate baseline.

\textbf{Network Topology Substantially Improves Forecasts When Local Case Histories Are Unavailable} We begin with models that do \emph{not} include town-level recent cases (\emph{non-AR} specifications). Relative to a Basic baseline that includes only macro-level incidence (statewide cases, MA), both network structure and mobility-volume interactions sharply increase out-of-sample performance (Table~\ref{tab:rotated_empirical_interaction_ar}). The one-week-ahead Predict-$R^2$ rises from \textbf{0.5977} (Basic) to \textbf{0.6616} when adding \emph{Betweenness} (Bet) alone and to \textbf{0.7055} when adding \emph{Intra-town flow} (Intra). The largest gains, however, come from interaction terms that tie macro incidence to network position or local mixing: \textbf{MA$\times$Bet} achieves \textbf{0.8290} and \textbf{MA$\times$Intra} achieves \textbf{0.8909}. The \emph{Full interaction} model (both interactions) is essentially identical at \textbf{0.8924}. All models include town fixed effects.

\begin{table}[tb]
\centering
\caption{\textbf{Empirical data} $\;|\;$ Interaction and Autoregressive Model Comparisons. The target variable is the next week's town-level COVID cases.}
\label{tab:rotated_empirical_interaction_ar}

%
\begin{minipage}{\linewidth}

\begin{minipage}[t]{\linewidth}
\centering
{\footnotesize \textbf{No town case data} $\;|\;$ Panel A: Interaction Model Comparison}\\[2pt]
\begin{threeparttable}
    
    {\scriptsize\setlength{\tabcolsep}{3pt}
    \renewcommand{\arraystretch}{1.0}
    \begin{tabular}{@{}lcccccc@{}}
    \hline
    \hline  & \multicolumn{6}{c}{} \\
     & {\shortstack{Basic}}
     & {\shortstack{Bet}}
     & {\shortstack{Intra}}
     & {\shortstack{Bet +\\MA:Bet}}
     & {\shortstack{Intra +\\MA:Intra}} 
     & {\shortstack{Full\\Interaction}} \\
    \\[-1.8ex] & (1) & (2) & (3) & (4) & (5) & (6)\\
    \hline \\[-1.8ex]
      MA &  0.003$^{***}$ &  0.003$^{***}$ &  0.002$^{***}$ &  $-$0.0002$^{***}$ &  $-$0.001$^{***}$ &  $-$0.001$^{***}$ \\ 
      &  (0.0001) &  (0.0001) &  (0.0001) &  (0.0001) &  (0.00004) &  (0.00004) \\ 
      Bet &  &  26,361.040$^{***}$ &  &  $-$5,396.817$^{***}$ &  &  1,798.626$^{***}$ \\ 
      &  &  (639.764) &  &  (577.593) &  &  (458.042) \\ 
      Intra &  &  &  $-$0.030$^{***}$ &  &  $-$0.009$^{***}$ &  $-$0.010$^{***}$ \\ 
      &  &  &  (0.0005) &  &  (0.0003) &  (0.0004) \\ 
      MA:Bet &  &  &  &  1.409$^{***}$ &  &  $-$0.299$^{***}$ \\ 
      &  &  &  &  (0.014) &  &  (0.021) \\ 
      MA:Intra &  &  &  &  &  0.00000$^{***}$ &  0.00000$^{***}$ \\ 
      &  &  &  &  &  (0.000) &  (0.000) \\ 
      Constant &  $-$2.950 &  $-$27.195$^{**}$ &  151.159$^{***}$ &  17.879$^{**}$ &  41.107$^{***}$ &  41.225$^{***}$ \\ 
      &  (11.661) &  (10.981) &  (10.597) &  (8.277) &  (6.587) &  (6.589) \\ 
    \hline \\[-1.8ex] 
     Predict-$R^2$ &  0.5977 &  0.6616 &  0.7055 &  0.8290 &  0.8909 &  \textbf{0.8924} \\ 
     Observations &  13,271 &  13,271 &  13,271 &  13,271 &  13,271 &  13,271 \\ 
     R$^{2}$ &  0.589 &  0.636 &  0.680 &  0.794 &  0.878 &  0.880 \\ 
     Adjusted R$^{2}$ &  0.579 &  0.627 &  0.672 &  0.789 &  0.875 &  0.877 \\ 
    \hline
    \hline
    \end{tabular}}
    \begin{tablenotes}
        \scriptsize
        \item Bet: Betweenness Centrality (Best Holistic Measure), \\ Intra: Intra-town Mobility (Best Local Measure). \\
        All Models include Town Fixed Effect. $^{*}$p$<$0.1; $^{**}$p$<$0.05; $^{***}$p$<$0.01
    \end{tablenotes}
\end{threeparttable}
\end{minipage}
\par\medskip
\begin{minipage}[t]{\linewidth}
\centering
{\footnotesize \textbf{With town case data} $\;|\;$ Panel B: Autoregressive Model Comparison}\\[2pt]
\begin{threeparttable}
    
    {\scriptsize\setlength{\tabcolsep}{3pt}
    \renewcommand{\arraystretch}{1.0}
    \begin{tabular}{@{}lcccccccc@{}}
    \hline 
    \hline  & \multicolumn{8}{c}{} \\  
     & {\shortstack{AR\\Baseline}} 
     & {\shortstack{AR +\\MA}} 
     & {\shortstack{AR +\\Bet}} 
     & {\shortstack{AR +\\Intra}} 
     & {\shortstack{AR + MA\\+ Bet}} 
     & {\shortstack{AR + MA\\+ Intra}} 
     & {\shortstack{AR + MA\\+ Bet + Int}} 
     & {\shortstack{AR + MA\\+ Intra + Int}} \\
    \\[-1.8ex] & (1) & (2) & (3) & (4) & (5) & (6) & (7) & (8)\\ 
    \hline \\[-1.8ex] 
       $town_{t}$ &  0.924$^{***}$ &  0.920$^{***}$ &  0.938$^{***}$ &  0.917$^{***}$ &  0.936$^{***}$ &  0.912$^{***}$ &  0.897$^{***}$ &  0.761$^{***}$ \\ 
      &  (0.004) &  (0.004) &  (0.004) &  (0.004) &  (0.004) &  (0.005) &  (0.006) &  (0.009) \\ 
       MA &  &  0.0001$^{**}$ &  &  &  0.00004 &  0.0001$^{**}$ &  $-$0.0001$^{**}$ &  $-$0.0002$^{***}$ \\ 
      &  &  (0.00003) &  &  &  (0.00003) &  (0.00003) &  (0.00003) &  (0.00003) \\ 
      Bet &  &  &  $-$2,905.778$^{***}$ &  &  $-$2,861.719$^{***}$ &  &  $-$4,098.917$^{***}$ &  \\ 
      &  &  &  (327.496) &  &  (329.315) &  &  (358.748) &  \\ 
      Intra &  &  &  &  $-$0.001$^{***}$ &  &  $-$0.001$^{***}$ &  &  $-$0.001$^{***}$ \\ 
      &  &  &  &  (0.0003) &  &  (0.0003) &  &  (0.0003) \\ 
      MA:Bet &  &  &  &  &  &  &  0.108$^{***}$ &  \\ 
      &  &  &  &  &  &  &  (0.013) &  \\ 
      MA:Intra &  &  &  &  &  &  &  &  0.00000$^{***}$ \\ 
      &  &  &  &  &  &  &  &  (0.000) \\ 
      Constant &  2.760 &  2.232 &  5.298 &  7.069 &  4.952 &  6.834 &  7.091 &  7.349 \\ 
      &  (5.149) &  (5.153) &  (5.141) &  (5.357) &  (5.148) &  (5.357) &  (5.140) &  (5.277) \\ 
    \hline \\[-1.8ex] 
     Predict-$R^2$ &  0.8986 &  0.8988 &  0.8977 &  0.8992 &  0.8979 &  0.8995 &  0.8991 &  \textbf{0.9041} \\ 
     Observations &  13,271 &  13,271 &  13,271 &  13,271 &  13,271 &  13,271 &  13,271 &  13,271 \\ 
     R$^{2}$ &  0.920 &  0.920 &  0.920 &  0.920 &  0.920 &  0.920 &  0.921 &  0.922 \\ 
     Adjusted R$^{2}$ &  0.918 &  0.918 &  0.918 &  0.918 &  0.918 &  0.918 &  0.919 &  0.920 \\ 
    \hline 
    \hline
    \end{tabular}}
    \begin{tablenotes}
        \scriptsize
        \item AR: autoregressive term: number of cases at the town-level in the current week. \\
        All Models include Town Fixed Effect. $^{*}$p$<$0.1; $^{**}$p$<$0.05; $^{***}$p$<$0.01
    \end{tablenotes}
\end{threeparttable}
\end{minipage}

\end{minipage}%

\end{table}

\textbf{With Town-Level Histories, Autoregression Dominates} When town-level recent cases are available (\emph{AR} specifications), autoregression captures most near-term predictability (Table~\ref{tab:rotated_empirical_interaction_ar} (Panel B)). The AR baseline reaches a one-week-ahead Predict-$R^2$ of \textbf{0.8986}. Augmenting with macro incidence, betweenness, intra-town flow, and their interactions yields only small increments; the best specification (\textbf{AR + MA + Intra + Int}) reaches \textbf{0.9041}, a gain of roughly \textbf{+0.55 percentage points} over the AR baseline.

This establishes a conditional result: \emph{topology is highly useful when granular histories are unavailable, but its marginal value is small when those histories are present}.

This empirical contrast is consistent with a simple decomposition: the partial-$R^2$ of the interaction regressor $M_t Z_{i,t}$ depends on $M_t^2$ and the cross-sectional variance of network exposure. Without town histories in the baseline, this term captures nearly all cross-town predictability; once recent cases are included, the usable variance of $Z_{i,t}$ is reduced after orthogonalization with respect to $y_{i,t}$, and the incremental explanatory power is correspondingly small.

\textbf{When Do Network Features Help Most? Epidemic Phases and Timing} We next compare performance across epidemic phases using the same model families. Over the \emph{Whole period}, Basic vs.\ interaction models reproduces the non-AR pattern (Predict-$R^2$ 0.5977 $\rightarrow$ 0.8924); restricting to \emph{wave periods} further increases absolute performance (Two Waves: 0.7502 $\rightarrow$ 0.9170). Comparing \emph{Two Waves} to the \emph{Smooth} period shows the same qualitative pattern with larger deltas during waves (Two Waves: Basic 0.7502 vs.\ Combined 0.9170; Smooth: Basic 0.7299 vs.\ Combined 0.8806). Splitting wave periods by timing shows that improvements are large in both waves but especially pronounced early: in the \emph{First wave}, Basic 0.5328 rises to 0.8663 (Combined), while in the \emph{Second wave}, the Basic model is already stronger (0.8125) and the Combined model reaches 0.9201.

These phase patterns align with two complementary robustness checks reported in the Supplementary Material. First, weekly Moran's $I$ statistics computed on population-normalized case rates indicate nontrivial spatial clustering that intensifies during surge periods, consistent with the idea that inter-town coupling becomes more informative when transmission is heterogeneous in space \cite{Moran_1950}. Second, spatial panel specifications that account for spatial dependence yield qualitatively consistent evidence that prevalence-gated network effects remain salient when spatial autocorrelation is explicitly modeled (see Supplementary Material, Spatial Analysis).

\begin{figure}[!t]
\centering
\includegraphics[width=\textwidth]{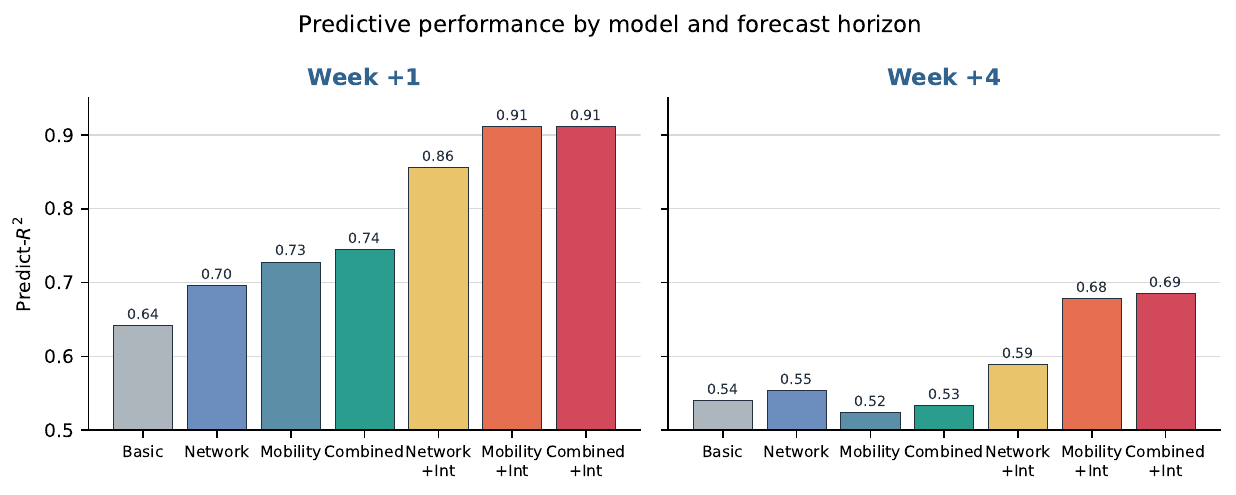}
\caption{Prediction Horizons. Model performance (Predictive $R^2$) for one-week and four-week ahead forecasts of the town-level case numbers across seven model specifications. The Basic model uses only macro-level incidence (statewide cases). The combined model uses both holistic Network (Betweenness) and local Mobility (Intra-weight) metrics. All models show performance degradation from week +1 to week +4. Models incorporating interaction terms between the mobility/network metric and the statewide cases (+Int) consistently achieve the highest $R^2$ values at both time horizons.}
\label{fig:prediction_horizons}
\end{figure}

\textbf{Forecast Horizon: Relative Advantages Persist Beyond One Week} Although absolute accuracy decays with forecast horizon, the relative
advantage of interaction models persists (Fig.~\ref{fig:prediction_horizons}).
At one week ahead, MA$\times$Intra achieves Predict-$R^2 = 0.891$
versus $0.598$ for Basic; at four weeks ahead, the gap narrows but
remains substantial ($0.741$ vs.\ $0.539$). MA$\times$Bet shows a
similar pattern ($0.829 \rightarrow 0.690$). This robustness across
horizons suggests that prevalence-gated mobility interactions retain
operational value for multi-week planning cycles, not only for
one-step-ahead forecasts.

\section{Generalizability via Agent-Based Simulation}\label{sec:generalizability}

The results above come from a single state and pathogen. To test whether the substitution--complementarity boundary reflects a \emph{generic} mechanism rather than idiosyncrasies of the Massachusetts data, we reproduce the analysis in a controlled, deliberately uncalibrated agent-based model (ABM) and probe its robustness across network structures, topological heterogeneity, endogenous mobility, and primary-channel quality. Full model details and parameters are in the Supplementary Material.

\textbf{Agent-Based Model (ABM)} To provide model-based validation of the analytical mechanism, we implemented an agent-based simulation model that simulates epidemics on \emph{generic} synthetic networks with known structure, deliberately \emph{uncalibrated} to the observed Massachusetts mobility system, since the absence of calibration is precisely what gives the test its force. The ABM provides a controlled environment in which to test whether the substitution–complementarity pattern identified in real data arises from a general mechanism (macro-scale prevalence modulating the predictive salience of network position) or from artifacts of measurement noise, reporting idiosyncrasies, or specification choices inherent in observational analysis. Unlike the reduced-form regressions, which rely on imperfect case reports and aggregated mobility metrics, the ABM offers full observability: we can directly track infections, movements, and changes in network structure under known transmission rules and controlled parameter variation. This allows us to systematically vary macro-level prevalence, network configurations, and the presence or absence of local case histories, generating synthetic case trajectories against which we run the same forecasting models as in the empirical analysis. If the regime-dependent pattern is specific to the Massachusetts data, the ABM would fail to reproduce it; if it reflects a generic property of network-mediated propagation under varying information conditions, the ABM should recapitulate the same conditional results.

In the model, each town is represented as a node in a directed, weighted network; edge weights capture inter-town flows, and self-loops capture within-town mixing. To capture the heterogeneity of human contact patterns, we generate the base connectivity using a Barab\'{a}si--Albert scale-free network, which is widely regarded as a good approximation of real social interaction structures \cite{barabasi1999emergence,pastor2001epidemic,cao2022micro}. Agents follow a stochastic SIR process, progressing through infection and recovery with probabilities tuned to epidemiological parameters. Movement occurs along edges in proportion to flow weights, coupling local epidemic dynamics to the evolving connectivity structure. Full parameter values, network families, initialization, and update rules are provided in the Supplementary Material.

\textbf{Mechanistic triangulation.} Running the identical forecasting battery on the synthetic data recovers the empirical pattern (Table~\ref{tab:rotated_abm_interaction_ar}). In the \emph{non-AR} ABM, the Basic model's Predict-$R^2$ is 0.5513; \textbf{MA$\times$Bet} reaches 0.8799, mirroring the empirical finding that interaction terms dominate when local histories are absent. In the \emph{AR} ABM (Table~\ref{tab:rotated_abm_interaction_ar} (Panel B)), the AR baseline is already strongest at 0.8653; the addition of network measures and their interactions does not enhance the prediction power.

Together, the ABM results reinforce a mechanistic interpretation in which \emph{macro-scale prevalence modulates the predictive salience of network topology}, and where local case histories, when available, largely subsume near-term network signal.

\begin{table}[tb]
\centering
\caption{\textbf{ABM data} $\;|\;$ Interaction and Autoregressive Model Comparisons. The target variable is the next week's town-level COVID cases.}
\label{tab:rotated_abm_interaction_ar}

%
\begin{minipage}{\linewidth}

\begin{minipage}[t]{\linewidth}
\centering
{\footnotesize \textbf{No town case data} $\;|\;$ Panel A: Interaction Model Comparison}\\[2pt]
\begin{threeparttable}
    
    {\scriptsize\setlength{\tabcolsep}{3pt}
    \renewcommand{\arraystretch}{1.0}
    \begin{tabular}{@{}lcccccc@{}}
    \hline
    \hline  & \multicolumn{6}{c}{} \\
     & {\shortstack{Basic}}
     & {\shortstack{Bet}}
     & {\shortstack{Intra}}
     & {\shortstack{Bet +\\MA:Bet}}
     & {\shortstack{Intra +\\MA:Intra}}
     & {\shortstack{Full\\Interaction}} \\
    \\[-1.8ex] & (1) & (2) & (3) & (4) & (5) & (6)\\
    \hline \\[-1.8ex]
      MA &  0.0054$^{***}$ &  0.0056$^{***}$ &  0.00087$^{***}$ &  0.00059$^{***}$ &  0.0058$^{***}$ &  0.0059$^{***}$ \\
      &  (0.00016) &  (0.00016) &  (0.00017) &  (0.00016) &  (0.00069) &  (0.00058) \\
      Bet &  &  $-$7,743$^{***}$ &  &  2,337$^{**}$ &  &  6,588$^{***}$ \\
      &  &  (1594) &  &  (1066) &  &  (1039) \\
      Intra &  &  &  0.056$^{***}$ &  &  0.059$^{***}$ &  0.027$^{***}$ \\
      &  &  &  (0.00156) &  &  (0.00159) &  (0.00224) \\
      MA:Bet &  &  &  &  1.133$^{***}$ &  &  0.838$^{***}$ \\
      &  &  &  &  (0.0261) &  &  (0.0387) \\
      MA:Intra &  &  &  &  &  $-$0.00000$^{***}$ &  $-$0.00000$^{***}$ \\
      &  &  &  &  &  (0.000) &  (0.000) \\
      Constant &  $-$2.371 &  11.8$^{**}$ &  $-$398$^{***}$ &  $-$4.110 &  $-$420$^{***}$ &  $-$200$^{***}$ \\
      &  (3.890) &  (4.841) &  (11.345) &  (3.180) &  (11.530) &  (16.634) \\
    \hline \\[-1.8ex]
     Predict-$R^2$ &  0.5513 &  0.5777 &  0.8286 &  \textbf{0.8799} &  0.8316 &  0.8623 \\
     Observations &  1,600 &  1,600 &  1,600 &  1,600 &  1,600 &  1,600 \\
     R$^{2}$ &  0.588 &  0.595 &  0.786 &  0.828 &  0.794 &  0.851 \\
     Adjusted R$^{2}$ &  0.529 &  0.537 &  0.755 &  0.803 &  0.764 &  0.829 \\
    \hline
    \hline
    \end{tabular}}
    \begin{tablenotes}
        \scriptsize
        \item Bet: Betweenness Centrality (Best Holistic Measure), \\ Intra: Intra-town Mobility (Best Local Measure). \\
        All Models include Town Fixed Effect. $^{*}$p$<$0.1; $^{**}$p$<$0.05; $^{***}$p$<$0.01
    \end{tablenotes}
\end{threeparttable}
\end{minipage}
\par\medskip
\begin{minipage}[t]{\linewidth}
\centering
{\footnotesize \textbf{With town case data} $\;|\;$ Panel B: Autoregressive Model Comparison}\\[2pt]
\begin{threeparttable}
    
    {\scriptsize\setlength{\tabcolsep}{3pt}
    \renewcommand{\arraystretch}{1.0}
    \begin{tabular}{@{}lcccccccc@{}}
    \hline
    \hline  & \multicolumn{8}{c}{} \\
     & {\shortstack{AR\\Baseline}}
     & {\shortstack{AR +\\MA}}
     & {\shortstack{AR +\\Bet}}
     & {\shortstack{AR +\\Intra}}
     & {\shortstack{AR + MA\\+ Bet}}
     & {\shortstack{AR + MA\\+ Intra}}
     & {\shortstack{AR + MA\\+ Bet + Int}}
     & {\shortstack{AR + MA\\+ Intra + Int}} \\
    \\[-1.8ex] & (1) & (2) & (3) & (4) & (5) & (6) & (7) & (8)\\
    \hline \\[-1.8ex]
       $town_{t}$ &  2.262$^{***}$ &  2.375$^{***}$ &  2.270$^{***}$ &  2.561$^{***}$ &  2.369$^{***}$ &  2.626$^{***}$ &  2.186$^{***}$ &  2.626$^{***}$ \\
      &  (0.0252) &  (0.0386) &  (0.0254) &  (0.0695) &  (0.0389) &  (0.0721) &  (0.0781) &  (0.0694) \\
       MA &  &  $-$0.00048$^{***}$ &  &  &  $-$0.00044$^{***}$ &  $-$0.00041$^{***}$ &  $-$0.00047$^{***}$ &  0.0045$^{***}$ \\
      &  &  (0.0001) &  &  &  (0.0001) &  (0.0001) &  (0.0001) &  (0.0005) \\
      Bet &  &  &  $-$1,770$^{**}$ &  &  $-$954 &  &  $-$469 &  \\
      &  &  &  (809) &  &  (842) &  &  (859) &  \\
      Intra &  &  &  &  $-$0.0096$^{***}$ &  &  $-$0.0086$^{***}$ &  &  $-$0.0055$^{***}$ \\
      &  &  &  &  (0.002) &  &  (0.002) &  &  (0.002) \\
      MA:Bet &  &  &  &  &  &  &  0.113$^{***}$ &  \\
      &  &  &  &  &  &  &  (0.0420) &  \\
      MA:Intra &  &  &  &  &  &  &  &  $-$0.00000$^{***}$ \\
      &  &  &  &  &  &  &  &  (0.000) \\
      Constant &  0.679 &  1.061 &  3.972 &  68.7$^{***}$ &  2.801 &  62.1$^{***}$ &  1.902 &  40.0$^{***}$ \\
      &  (2.029) &  (2.022) &  (2.524) &  (14.930) &  (2.539) &  (15.019) &  (2.555) &  (14.611) \\
    \hline \\[-1.8ex]
     Predict-$R^2$ &  \textbf{0.8653} &  0.8517 &  0.8618 &  0.8589 &  0.8509 &  0.8474 &  0.8524 &  0.8212 \\
     Observations &  1,600 &  1,600 &  1,600 &  1,600 &  1,600 &  1,600 &  1,600 &  1,600 \\
     R$^{2}$ &  0.888 &  0.889 &  0.888 &  0.889 &  0.889 &  0.890 &  0.890 &  0.898 \\
     Adjusted R$^{2}$ &  0.872 &  0.873 &  0.872 &  0.873 &  0.873 &  0.874 &  0.873 &  0.884 \\
    \hline
    \hline
    \end{tabular}}
    \begin{tablenotes}
        \scriptsize
        \item AR: autoregressive term: number of cases at the town-level in the current week. \\
        All Models include Town Fixed Effect. $^{*}$p$<$0.1; $^{**}$p$<$0.05; $^{***}$p$<$0.01
    \end{tablenotes}
\end{threeparttable}
\end{minipage}

\end{minipage}%

\end{table}

\begin{figure}[!t]
\centering
\includegraphics[width=\textwidth]{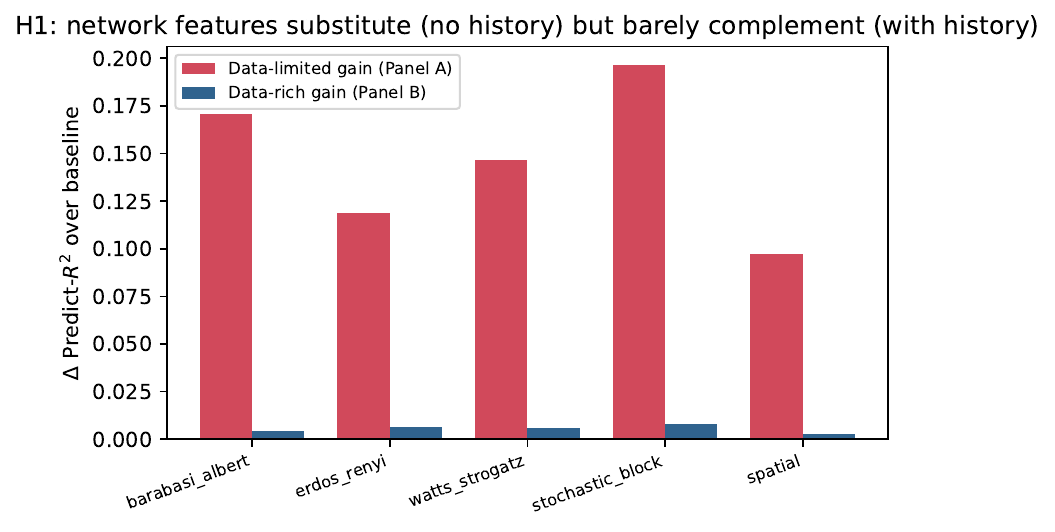}
\caption{\textbf{H1: the regime flip is generic across network families.} For each synthetic network family, the data-limited gain (Panel~A: best macro$\times$network model minus the macro-only baseline) is large, while the data-rich gain (Panel~B: best AR$+$network model minus the AR baseline) is near zero. Network features substitute for missing local histories but barely complement an informative one, regardless of topology.}
\label{fig:abm_h1}
\end{figure}

\textbf{The regime flip is generic across network families (H1).} We first ask whether the boundary survives when the network is generated by mechanisms other than the scale-free topology of the base case. Across five families (Barab\'asi--Albert, Erd\H{o}s--R\'enyi, Watts--Strogatz, stochastic-block, and spatial random-geometric) the qualitative flip holds in every case: without town-level histories the macro$\times$network interaction substantially beats the macro-only baseline (data-limited Predict-$R^2$ gains of roughly $+0.1$ to $+0.2$), whereas with an autoregressive baseline the same network features add essentially nothing ($\le +0.01$; Fig.~\ref{fig:abm_h1}). The boundary is therefore not an artifact of any particular topology.

\begin{figure}[!t]
\centering
\includegraphics[width=3.4in]{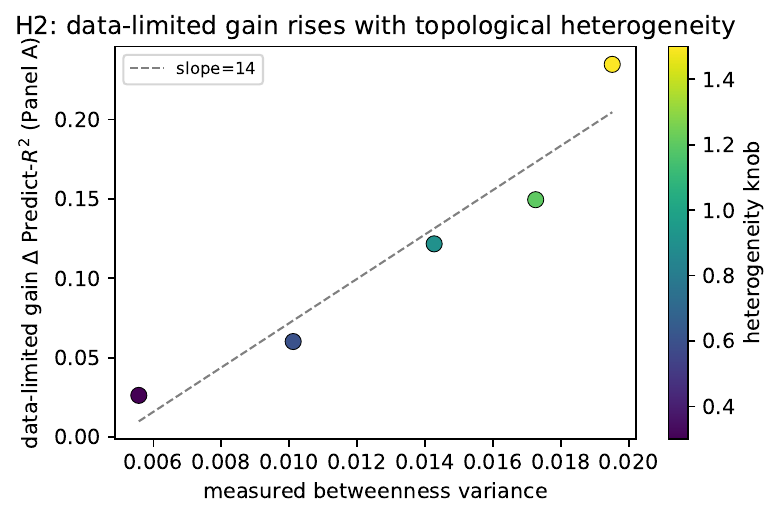}
\caption{\textbf{H2: data-limited gain rises with topological heterogeneity.} Across heterogeneity levels (colour), the data-limited Predict-$R^2$ gain increases with the measured betweenness variance of the network.}
\label{fig:abm_h2}
\end{figure}

\textbf{The data-limited gain grows with topological heterogeneity (H2).} Sweeping the network heterogeneity knob within a family, the data-limited gain rises monotonically with the \emph{measured} betweenness variance of the generated networks (Fig.~\ref{fig:abm_h2}); at the lowest heterogeneity the flip does not register at all. This makes the mechanism explicit: network position carries predictive value precisely when positions differ enough to matter, consistent with the constructive derivation in which the structural multiplier $Z_{j,t}$ enters the interaction term.

\begin{figure}[!t]
\centering
\includegraphics[width=\textwidth]{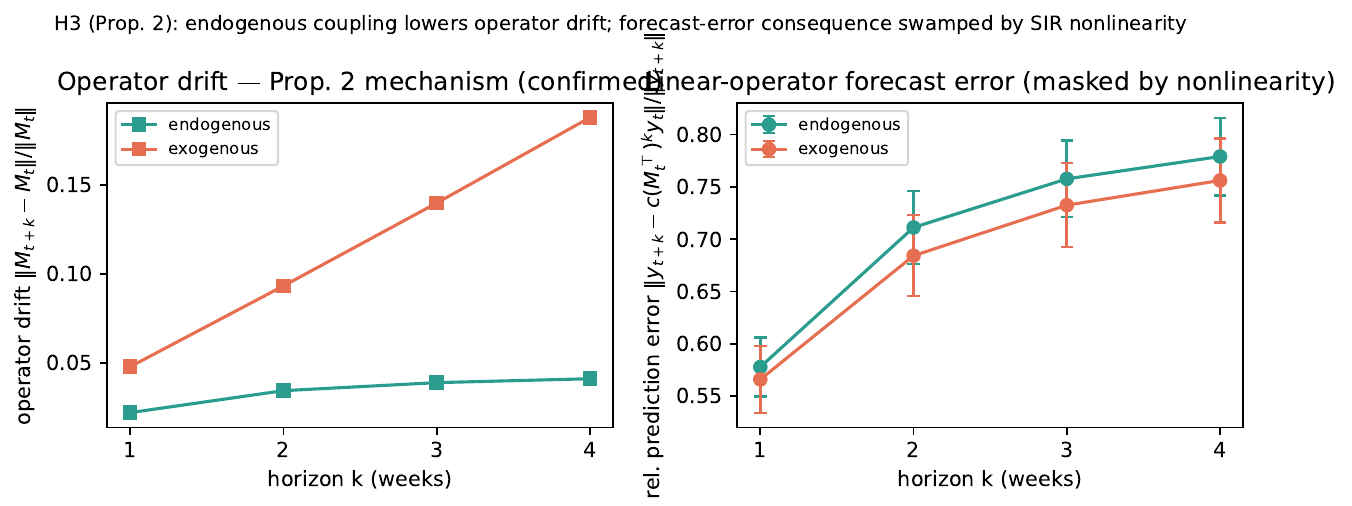}
\caption{\textbf{H3 / Proposition~\ref{prop:multistep}: endogenous coupling lowers operator drift.} Left: operator drift grows much faster under exogenous than endogenous mobility. Right: the iterated linear-operator forecast error does not separate the two, because nonlinear dynamics dominate the linearization error.}
\label{fig:abm_h3}
\end{figure}

\textbf{Endogenous mobility constrains operator drift (H3, Proposition~\ref{prop:multistep}).} To probe Proposition~\ref{prop:multistep} we contrast endogenous mobility (where the weekly interaction matrix responds to prevalence) against a variance- and smoothness-matched exogenous null in which the same mobility variation is decoupled from the epidemic. The operator drift $\|M_{t+k}-M_t\|/\|M_t\|$ is markedly lower under endogenous coupling, and the gap widens with horizon (about a factor of four by four weeks; Fig.~\ref{fig:abm_h3}, left), exactly the quantity by which Proposition~\ref{prop:multistep} bounds multi-step bias. The downstream forecast-error consequence is, however, masked in finite samples: because the realized dynamics are nonlinear, the iterated \emph{linear} operator predictor carries substantial linearization error that swamps the drift-driven term (Fig.~\ref{fig:abm_h3}, right). The proposition's \emph{mechanism} is thus confirmed in simulation, while its forecast-bias prediction is a property of the linearization.

\begin{figure}[!t]
\centering
\includegraphics[width=\textwidth]{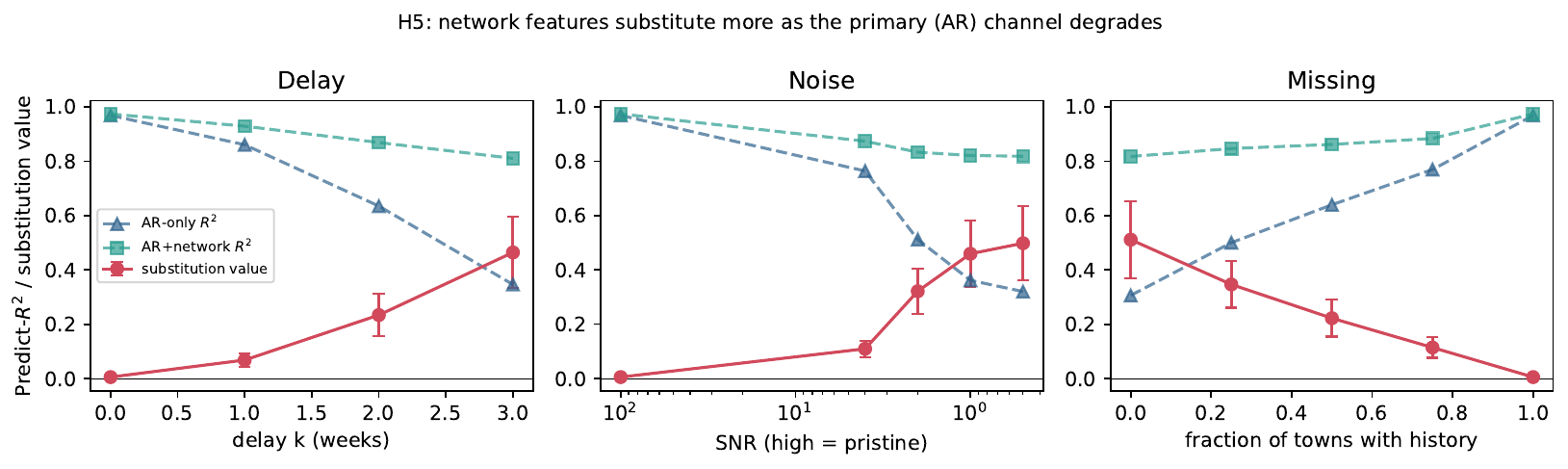}
\caption{\textbf{H5: substitution value vs.\ primary-channel degradation.} As the AR channel is degraded by delay, noise, or missingness, the AR-only skill (blue) falls, the AR$+$network skill (green) is preserved, and the substitution value (red) rises from $\approx 0$ to a large lift ($\approx 0.5$).}
\label{fig:abm_h5}
\end{figure}

\textbf{Substitution value grows as the primary channel degrades (H5).} Finally, we degrade the autoregressive (primary-surveillance) channel three independent ways (delay, measurement noise, and missingness) and track the network lift over the AR baseline. In all three, the substitution value rises smoothly and monotonically from near zero, when the channel is pristine, to a large lift ($\approx 0.5$ Predict-$R^2$) when the channel is fully degraded (Fig.~\ref{fig:abm_h5}). This turns the binary regime distinction into a continuous curve and quantifies the design rule: the value of integrating the behavioral signal scales with how degraded the primary channel is.

\begin{figure}[!htbp]
\centering
\includegraphics[width=3.4in]{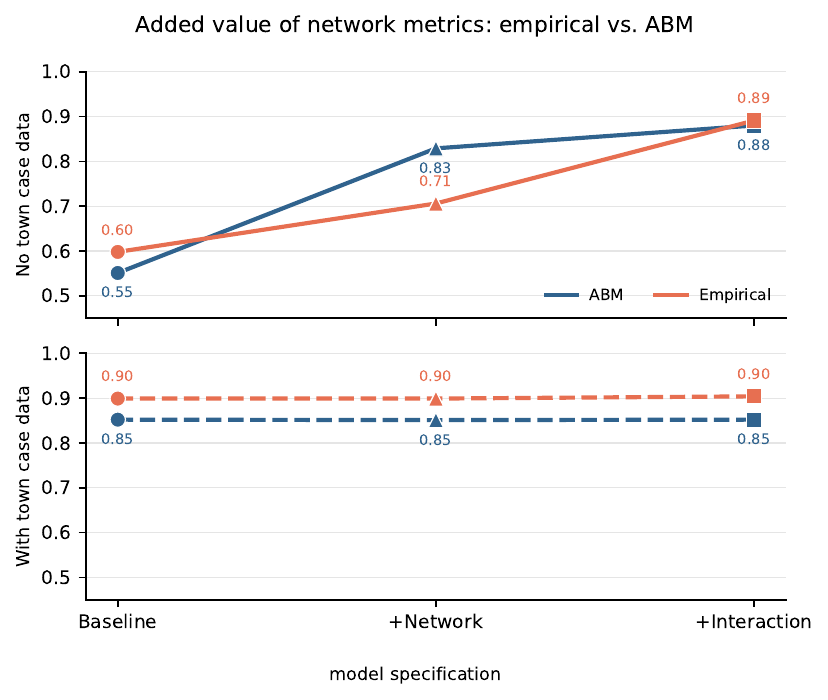}
\caption{Added predictive value of network metrics and their interaction with state-level information, shown as a slopegraph of out-of-sample Predict-$R^2$. Points denote model variants for both the Empirical and ABM scenarios. Each trend compares model performance under No town data (top) and With town data (bottom). The results show that when high-resolution town-level data are available, incorporating network metrics (and their interactions) provides little to no additional predictive benefit.}
\label{fig:overall_r2_comparison}
\end{figure}

\textbf{Summary} Across data and model families, we find a consistent conditional pattern. When granular town-level histories are \emph{unavailable}, interactions between macro incidence and mobility-network features deliver large out-of-sample gains (nearly 50\% increase in predictive power). When town-level histories are \emph{available}, autoregression dominates, and network features add a small, statistically significant but practically minimal increase (best AR model 0.9041 vs.\ baseline 0.8986). Gains are largest during waves and rising phases, and the same qualitative pattern holds in the ABM. The overall comparison is illustrated in Fig.~\ref{fig:overall_r2_comparison}. This conditional pattern (strong substitution, weak complementarity) constitutes the empirical basis for the system-design principles discussed in Section~\ref{sec:discussion}.

\section{Discussion}\label{sec:discussion}
This paper set out to answer a question that extends beyond epidemic modeling: under what conditions does a dynamic behavioral signal justify its integration cost in a sociotechnical monitoring system? The answer matters because sociotechnical systems exhibit endogenous behavioral response, the agents being monitored adapt their mixing, movement, and interaction in ways that feed back into the outcomes the system tracks, and the value of sensing this adaptation depends on what the primary monitoring channel already captures.

Using town-level COVID-19 forecasting in Massachusetts as an empirical laboratory, we found a clear substitution--complementarity boundary. When the primary surveillance channel is degraded (granular town-level case histories are missing, delayed, or coarsened) coupling macro-scale incidence with a place's position in the mobility network materially improves short-horizon forecasts (Predict-$R^2$: ${\sim}0.60 \rightarrow {\sim}0.83$--$0.89$). When the primary channel is intact (timely local histories are available) autoregressive baselines already capture most week-ahead signal, and the same mobility features offer only marginal complementarity (${\approx}{+}0.5$ percentage points; Table~\ref{tab:rotated_empirical_interaction_ar}). This regime-dependent pattern is consistent with the broader finding of Bjerre-Nielsen et al.\ \cite{BjerreNielsen2021} that task-specific outcome histories outperform surveillance-style behavioral data, and it refines long-standing insights from network epidemiology (that structure conditions transmission potential \cite{keeling2005networks,Barrat2004,Colizza2007,balcan2009multiscale,Brockmann2013}) by specifying \emph{when} that structural information is operationally valuable for prediction.

\textbf{Mechanism: Why Substitution Is Strong and Complementarity Is Weak} The analytical decomposition makes the regime boundary precise. In the macro-only baseline, the interaction term $MA_t \, Z_{i,t}$ accounts for the full $MA_t$-scaled cross-sectional variation in network exposure, producing a large partial-$R^2$. When recent town cases enter the baseline, much of this cross-sectional variation is already captured by autoregression: the Frisch--Waugh--Lovell orthogonalization reduces the usable variance of $Z_{i,t}$, and the residual explanatory contribution of network exposure is correspondingly small. This duality is not specific to epidemiology, it applies whenever an interaction between a common contextual variable and a cross-sectional behavioral feature competes with an autoregressive term that already embeds the consequences of recent behavior. The same logic would predict, for instance, that interbank transaction topology adds forecasting value for financial stress when balance-sheet data are stale, but not when real-time
position data are available \cite{acemoglu2015systemic,brunetti2019interconnectedness}.

Across empirical models, the largest improvements came from \emph{prevalence-gated interaction} terms that couple macro-scale incidence with network position (e.g., MA$\times$Bet; Table~\ref{tab:rotated_empirical_interaction_ar}). This implies that the marginal effect of centrality strengthens as statewide incidence rises: when macro risk is high, towns that bridge communities or sit near flow bottlenecks are more informative for next-week spread than when macro risk is low. Performance improvements are larger during epidemic waves than during smoother phases, and are particularly pronounced early in the pandemic, consistent with periods in which contact patterns and inter-community connectivity shift rapidly \cite{hale2021global,chinazzi2020effect,prem2020effect}.

\textbf{How Much Behavioral Detail Is Needed?} A practical question for system designers is the granularity of behavioral encoding required. In our non-AR comparisons, the specification augmenting the baseline with network position interacted with macro incidence achieves Predict-$R^2 \approx 0.8290$, recovering most of the improvement observed when adding the strongest mobility-volume interaction (Predict-$R^2 \approx 0.8909$), with the combined model essentially identical ($\approx 0.8924$; Table~\ref{tab:rotated_empirical_interaction_ar}). Thus, a substantial share of forecasting gains arises from capturing how \emph{inter-town connectivity structure} modulates risk as macro incidence changes, with remaining gains closed by additional mobility-volume information. This suggests that even coarse network summaries (e.g., a single centrality measure interacted with a prevalence indicator) can capture much of the available signal, a finding relevant to settings where detailed flow data are expensive or privacy-constrained.

\textbf{External Validity and the Transferable Template} Our analysis is limited to one U.S.\ state and one pathogen over April 2020--April 2021, raising a natural question about generalizability. We address this concern through three complementary strategies.

First, the agent-based model reproduces the empirical pattern under controlled dynamics: without local histories, interaction models (especially macro incidence $\times$ betweenness) deliver the largest gains; with local histories, AR baselines dominate, and added topology does not help (Table~\ref{tab:rotated_abm_interaction_ar}). Because the ABM operates on synthetic networks with known transmission rules, this triangulation confirms that the regime-dependent pattern arises from a generic mechanism (macro-scale prevalence modulating the predictive salience of network position) rather than from specification choices or measurement idiosyncrasies of the Massachusetts data \cite{keeling2008,Aleta2020}.

Second, the analytical decomposition is entirely domain-agnostic. The Frisch--Waugh--Lovell result that interaction terms lose incremental explanatory power once an autoregressive term absorbs cross-sectional variance holds for any linear specification where a common contextual variable is interacted with a cross-sectional feature. This means the substitution--complementarity boundary we document should appear wherever a dynamic behavioral signal's cross-sectional variance overlaps substantially with what recent outcome histories capture, a condition likely met in supply-chain disruption forecasting \cite{Zhao2022SupplyResilience,brintrup2016understanding}, financial contagion monitoring \cite{acemoglu2015systemic,glasserman2015contagion}, and other network-mediated propagation processes.

Third, although forecast horizons decay in absolute accuracy, the
\emph{relative} advantage of interaction models persists to 2--4 weeks
(Fig.~\ref{fig:prediction_horizons}), suggesting robustness across
operational planning cycles.

\textbf{Implications for System Design and Deployment} Our findings translate into a design principle applicable to any sociotechnical monitoring system that fuses dynamic behavioral data with outcome-oriented surveillance: \emph{treat the behavioral signal as a conditional module whose activation depends on the quality of the primary channel and the dynamic state of the monitored process \cite{Veit2023SystemResilience}.} In value-of-information terms \cite{howard1966information}, the substitution gain quantified as the primary channel degrades (Fig.~\ref{fig:abm_h5}) is precisely the marginal value of the behavioral stream; the module should be activated whenever that value exceeds its acquisition, computation, and privacy cost, a threshold policy that places the rule in the predictive-to-prescriptive analytics pipeline \cite{bertsimas2020predictive,elmachtoub2022smart}.

Concretely, we recommend a two-mode operational workflow with explicit triggers.

\emph{Data-limited mode.} Activated when primary surveillance is delayed, coarsened, or missing, or when macro-level indicators signal a rapid-change phase (e.g., rising epidemic prevalence, emerging supply disruption, financial stress spike). In this mode, compute prevalence-gated interaction features that couple the behavioral signal's network topology with contextual state, and use these to prioritize resources (surveillance, testing, messaging) toward structurally important nodes (bridge towns, hub ports, systemically connected institutions). The large forecasting gains we observe in this regime justify the collection, computation, and privacy costs of the behavioral stream.

\emph{Data-rich mode.} Activated when primary surveillance is timely and the monitored process is in a stable or declining phase. In this mode, rely on autoregressive baselines for core forecasts and retain the behavioral signal as a \emph{diagnostic overlay}: use network features to identify structurally important nodes that may seed spillovers, to detect structural breaks in mixing patterns, and to flag emerging connectivity shifts that could precede the next rapid-change phase. The marginal forecasting gain from full behavioral integration is small in this regime, and the simpler pipeline reduces latency, maintenance burden, and interpretability costs.

This tiered logic also informs data governance. Because the largest gains arise in the data-limited regime, privacy-preserving aggregated behavioral summaries may offer high operational value precisely when traditional surveillance is most constrained. Conversely, when surveillance is strong, the marginal forecasting benefit of incorporating additional behavioral features is small, which can inform proportionality decisions about data collection, compute allocation, and data-sharing agreements.

\section{Conclusion}\label{sec:conclusion}
Computational social systems increasingly face the question of whether to incorporate dynamic behavioral data (mobility traces, transaction flows, communication patterns) into forecasting and monitoring pipelines. We show that the answer is conditional. Using town-level COVID-19 forecasting as a rigorous empirical template, we demonstrate a substitution--complementarity boundary: mobility-derived network features provide large predictive gains when the primary surveillance channel (local case histories) is degraded, but only marginal gains when it is intact. An agent-based model and an analytical decomposition confirm that this boundary arises from a generic mechanism (not from the specifics of one pathogen or one geography) in which macro-scale contextual state modulates the predictive salience of network position, and in which autoregressive outcome histories, when available, subsume much of the cross-sectional signal that behavioral features provide.

The practical implication is a deployable design rule: activate
behavioral-signal integration when the primary channel is weak, or the
system is in a rapid-change phase; deactivate it, retaining only
diagnostic overlays, when the primary channel is strong, and conditions are stable. This conditional-module logic applies beyond epidemiology, to any domain where network-structured behavioral flows are fused with outcome-oriented surveillance under varying data-quality regimes.

Several limitations warrant acknowledgment in this work. Mobility signals and reported cases are measured with error; smartphone-based traces may underrepresent specific populations, and case reports vary with testing intensity and reporting delays. Our models are reduced-form and emphasize out-of-sample prediction; they abstract from explicit transmission dynamics and behavior--policy feedbacks. We study one geography and time window; effects may differ with variant properties, vaccination coverage, or policy regimes. We focus on a small set of topology measures; other structural features (e.g., time-varying community structure, edge-direction asymmetries) may add value. Although we guard against overfitting with cross-validation and report Predict-$R^2$, model-selection uncertainty remains. To partially address concerns about metric arbitrariness and spatial confounding, the Supplementary Material reports (i)~an ablation-style comparison over a broader set of centrality and local mobility candidates, which identifies betweenness and intra-town flow as the most predictive representatives in their respective classes, and (ii)~spatial diagnostics and spatial panel robustness checks that support the regime-dependent conclusions under explicit spatial dependence.

Future work should test this template in the cross-domain settings motivating our framing (supply-chain disruption forecasting, financial contagion monitoring, and urban demand prediction) and extend the analytical framework to accommodate nonlinear models, dynamic community detection, and adaptive triggering rules that switch between operational modes in real time.


\bibliographystyle{IEEEtran}
\bibliography{mobility_refs}
\section*{Data Availability Statement }
This study draws on two data sources. Town-level weekly COVID-19 case counts are publicly available from the Massachusetts Department of Public Health. Inter-town mobility was derived from aggregated, de-identified smartphone origin–destination data provided by SafeGraph under a data-use agreement; the raw mobility data are governed by that agreement and cannot be redistributed by the authors. The processed, de-identified town-week panel and the derived network features required to reproduce the reported results, together with all analysis code, will be deposited in a public online repository upon acceptance, with a persistent link provided in the published article.

\end{document}


\def\spacingset#1{\renewcommand{\baselinestretch}%
    {#1}\small\normalsize} \spacingset{1}

\maketitle
\tableofcontents
\clearpage

\section{Data Preparation}

\begin{figure*}[!htbp]
    \centering
    \includegraphics[width=\textwidth]{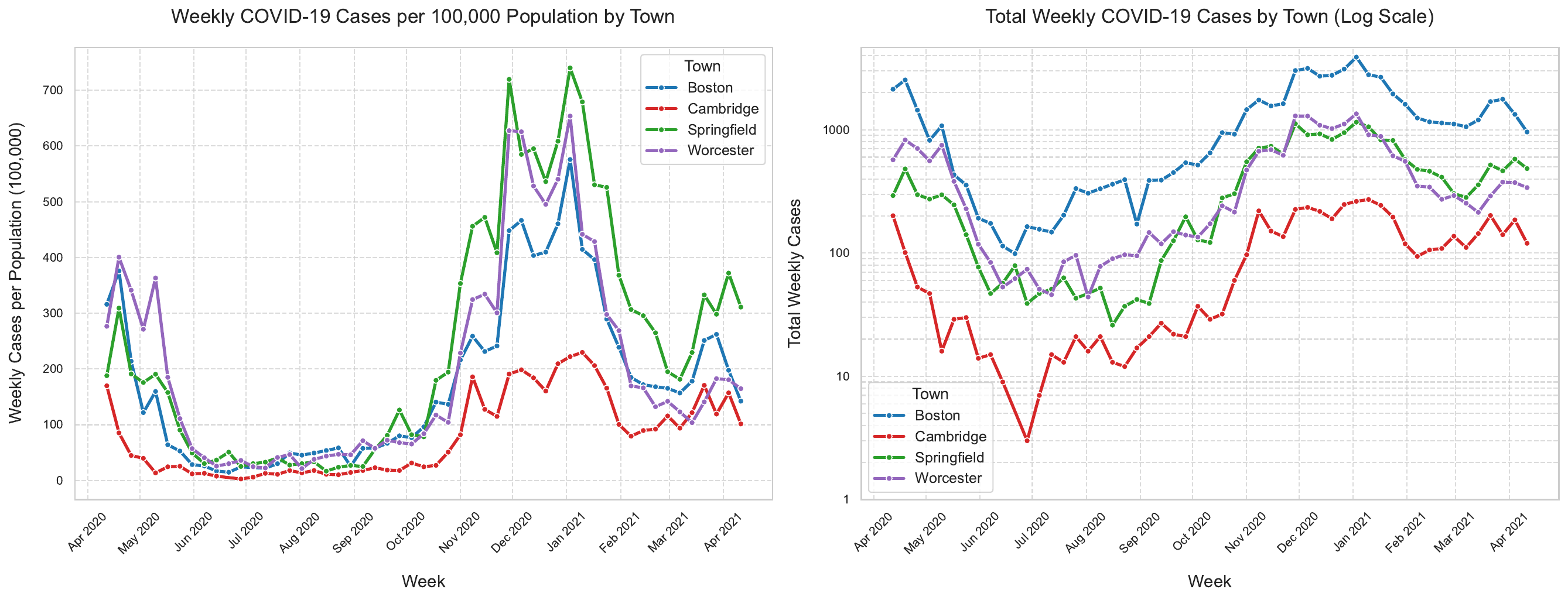}
    \caption{Comparison of weekly cases trend in major towns. Left: Absolute number of weekly cases. Right: Normalized number of weekly cases per 100,000 population}
    \label{fig:weekly_cases}
\end{figure*}

Figure \ref{fig:weekly_cases} visually illustrates the weekly COVID-19 case trends for four major Massachusetts cities (Boston, Cambridge, Springfield, and Worcester) spanning from April 2020 to April 2021. The left panel displays absolute weekly case counts, while the right panel normalizes these counts per 100,000 population. Notably, the normalized data reveal that Springfield experienced the highest infection rates relative to its population, reaching nearly 750 cases per 100,000 during its peak, highlighting the importance of population-adjusted metrics.

The observed patterns in Figure \ref{fig:weekly_cases} reflect the dynamic interplay of public health interventions, seasonal factors, and behavioral changes throughout the pandemic. The initial wave in spring 2020 corresponds to the early outbreak phase, characterized by limited testing and nascent mitigation strategies. The subdued case numbers during summer (June-September 2020) likely resulted from a combination of increased outdoor activities, stringent early lockdowns, and greater adherence to masking protocols. The dramatic surge during winter (November 2020 to January 2021) coincides with holiday gatherings, increased indoor activities due to colder weather, and potential pandemic fatigue. The subsequent decline from February 2021 onwards correlates with the rollout of vaccination efforts. Furthermore, the disproportionately high per-capita rates in Springfield may point to underlying socioeconomic disparities and a higher proportion of essential workers unable to work remotely, while Cambridge's consistently lower rates could be attributed to its demographic profile, a younger university-affiliated population, higher income levels potentially leading to greater compliance with preventive measures, and a robust public health infrastructure. These differentiated patterns underscore how the pandemic unevenly impacted communities along socioeconomic lines, even within the same state.

A sample of this aggregated weekly case data, showcasing both town-specific and statewide case counts, is presented in Table \ref{tab:case_data}. The table illustrates the weekly progression of confirmed COVID-19 cases for individual towns, alongside the corresponding total cases across Massachusetts, and the town-level cases for the subsequent week, which served as our prediction target.

\begin{table}[!htbp]
\centering
\caption{Sample of Weekly COVID-19 Confirmed Case Data for Massachusetts Towns.}
\label{tab:case_data}
\resizebox{0.75\textwidth}{!}{ 
\begin{tabular}{ccccc}
\toprule
\textbf{Town} & \textbf{Week End Date} & \textbf{Town Cases ($y_t$)} & \textbf{MA Cases ($MA_t$)} & \textbf{Town Cases ($y_{t+1}$)} \\
\midrule
Abington & 2020-04-12 & 24 & 15236 & 65 \\
Abington & 2020-04-19 & 65 & 14828 & 34 \\
Abington & 2020-04-26 & 34 & 12401 & 13 \\
Abington & 2020-05-03 & 13 & 9822 & 29 \\
Abington & 2020-05-10 & 29 & 7492 & 7 \\
Abington & 2020-05-17 & 7 & 5950 & 4 \\
\bottomrule
\end{tabular}}
\end{table}
\newpage

\newpage
\section{Methods}
\subsection{Building Town-level Network}\label{sec2}





To analyze mobility dynamics and their influence on pandemic spread at the town level, our initial step involved constructing a town-level mobility network from SafeGraph data. This dataset provides daily records of registered device movements between individual Census Block Groups (CBGs) within Massachusetts.

From this raw data, we first conceptualized the daily mobility pattern across all Massachusetts CBGs as a directed network. In this \textbf{daily CBG-level network}, each node represents a unique CBG. A directed edge from node A to node B signifies the movement of registered devices from CBG A to CBG B, with the edge weight representing the total number of such moving devices. To account for daily variations and establish a more stable representation of mobility, these daily CBG networks were aggregated into a \textbf{weekly CBG-level network}.

\begin{figure*}[!htbp]
  \centering
  \includegraphics[width=0.9\textwidth]{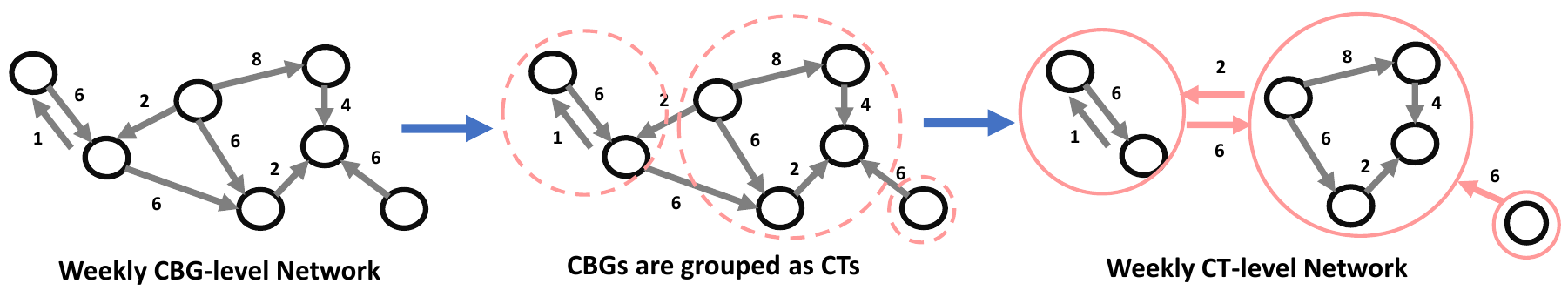}
  \caption{Network Generation. The diagram shows the conversion of the weekly CBG-level network to the weekly CT-level network.}
  \label{fig:cbg_to_ct}
\end{figure*}

Our next step involved transforming this weekly CBG-level network into a \textbf{weekly Census Tract (CT)-level network}. This was achieved by grouping CBG nodes and their corresponding edges. Specifically, all CBGs belonging to the same Census Tract were consolidated into a single CT node. The weight of an edge between two CT nodes was then calculated as the sum of all mobility flows from all CBGs within the origin CT to all CBGs within the destination CT. This hierarchical aggregation process is visually depicted in Figure \ref{fig:cbg_to_ct}.

Following the same hierarchical aggregation logic, the weekly CT-level network was further transformed into a \textbf{weekly town-level network}. All Census Tracts within a given town were grouped to form a town node, and the edges between towns represent the aggregate mobility flows between their constituent Census Tracts.

Through these systematic aggregation steps, we successfully created a comprehensive panel dataset. This dataset comprises over 16,500 records, providing weekly mobility network data for all 313 towns in Massachusetts, which subsequently formed the basis for calculating our network centrality measures and for the regression analyses discussed in the following sections.

\subsection{Calculating Network Centrality Measures}\label{sec3}

To characterize mobility patterns and network structure, we computed two categories of metrics for each node (town) in our network: (1) node-specific mobility metrics that quantify population movement about an individual town, and (2) network centrality metrics that assess each town's relative position and influence within the overall network topology. See SI Section 2 (Methods) for detailed info for all metrics.
These metrics collectively provide a comprehensive characterization of mobility dynamics and network topology, enabling the identification of critical nodes and movement patterns relevant to pandemic spread.

\subsubsection{Evolution of Network Measures over time}


\begin{figure}[!htbp]
    \centering
    \includegraphics[width=0.7\textwidth]{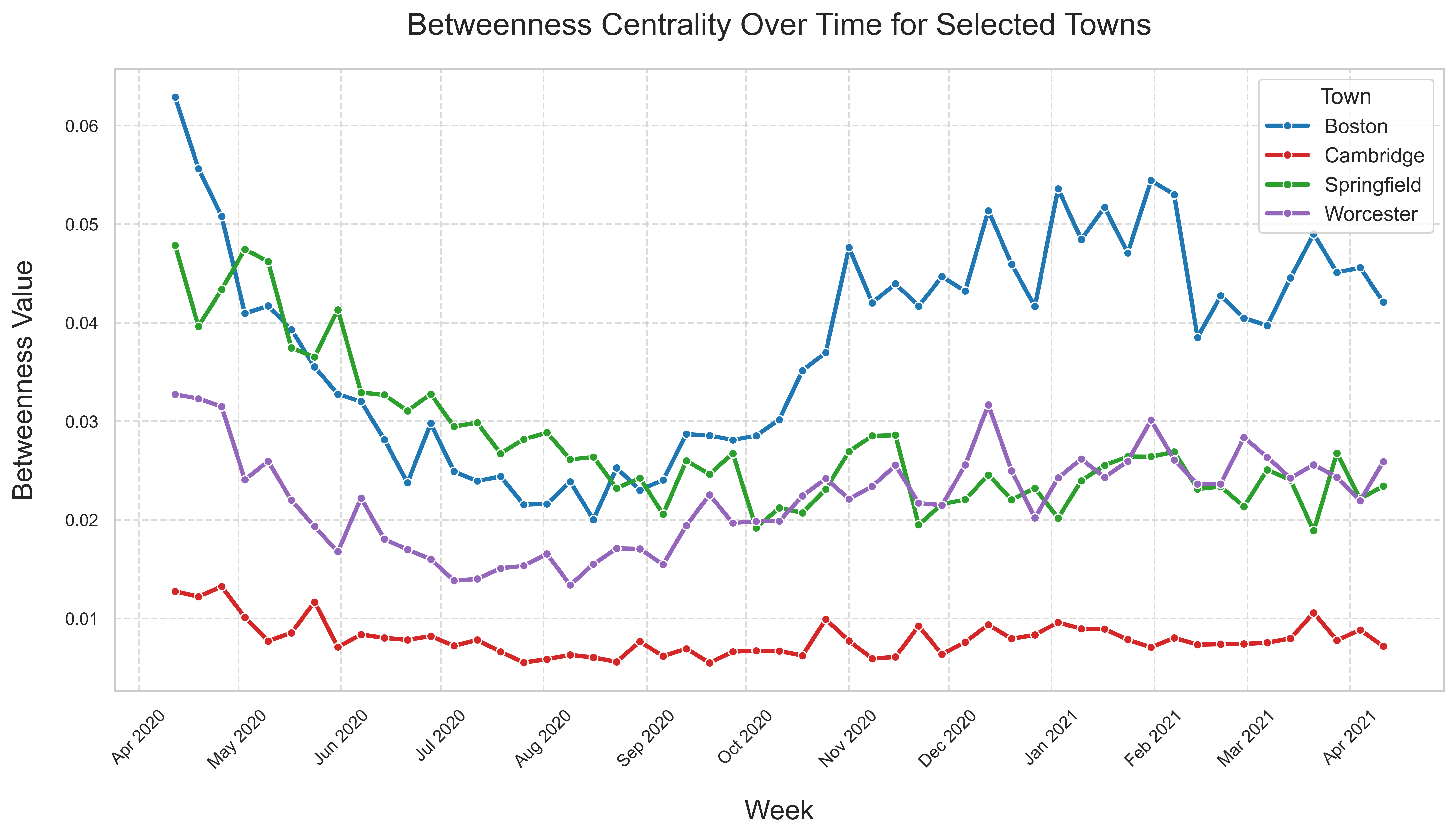}
    \caption{Betweenness centrality in major Massachusetts towns}
    \label{fig:betweenness_trend}
\end{figure}

Figure \ref{fig:betweenness_trend} illustrates the dynamic evolution of betweenness centrality for Massachusetts' four most populated towns throughout the study period, reflecting significant shifts in their roles within the state's mobility network. This temporal analysis of network measures provides crucial context for understanding their predictive power in our subsequent models.

Boston exhibits the most pronounced fluctuations. Its betweenness centrality started at the highest level in April 2020, consistent with its status as a primary hub. However, it experienced a sharp decline during the initial lockdown phase (April-June 2020), indicating a drastic reduction in its intermediary role as mobility was restricted. Centrality remained low through the summer but began a significant resurgence in October 2020 as restrictions eased, peaking in January-February 2021. This recovery aligns with observed peaks in weekly COVID-19 case patterns (as seen in Figure \ref{fig:weekly_cases}), suggesting Boston re-established itself as a critical connector in the state's mobility network as activity resumed.

In contrast, other major towns, such as Springfield, Worcester, and Cambridge, showed comparatively modest changes. Springfield maintained the second-highest initial centrality among these, declining more gradually than Boston. Cambridge consistently displayed the lowest betweenness centrality throughout the entire period, suggesting its less central role in broader inter-town mobility patterns, despite its population and economic significance. These observed temporal dynamics in network centrality are directly linked to the fluctuating mobility patterns that we hypothesize influence disease transmission, thereby underpinning the effectiveness of these metrics in our forecasting models presented in the next sections.

\subsection{Network Centrality Measures}

Table \ref{tab:network_metrics_info} presents the complete set of metrics utilized in our analysis. A sample of weekly network measures for towns is also presented in Table \ref{tab:network_measures}.

\begin{table*}[!htbp]
    \centering
    \small
    \caption{Network Metrics}
    \label{tab:network_metrics_info}
    \resizebox{\textwidth}{!}{
    \begin{tabular}{llll}
    \hline
    \textbf{Metric} & \textbf{Variable} & \textbf{Level} & \textbf{Description} \\
    \hline
    Incoming Flow & inweightt & Node & Total population influx from other towns \\
    Outgoing Flow & outweightt & Node & Total population outflux to other towns \\
    Intra-Town Flow & intraweightt & Node & Population retained within town boundaries (self-loops) \\
    Intra-Block Flow & intraweightb & Node & Population retained within Census Block Group \\
    Local Mobility Ratio (Town) & LMRt & Node & Proportion of mobility contained within town: $\frac{intraweightt}{intraweightt + outweightt}$ \\
    Relative LMR (Town) & RLMRt & Node & Town-level LMR normalized to cross-town average \\
    Local Mobility Ratio (Block) & LMRb & Node & Proportion of mobility contained within block \\
    Relative LMR (Block) & RLMRb & Node & Block-level LMR normalized to cross-town average \\
    \hline
    In-PageRank & in\_pr & Holistic & Town significance as destination in random walk models \\
    Out-PageRank & out\_pr & Holistic & Town significance as origin in random walk models \\
    Unweighted Betweenness & bet & Holistic & Bridge functionality in network (hop-based shortest paths) \\
    Weighted Betweenness & bet\_wei & Holistic & Bridge functionality on flow-efficient paths \\
    In-Closeness & in\_clo & Holistic & Network accessibility (minimum hops required from other towns) \\
    Out-Closeness & out\_clo & Holistic & Network reach (minimum hops required to other towns) \\
    Weighted In-Closeness & in\_clo\_wei & Holistic & Efficiency of accessibility via high-flow routes \\
    Weighted Out-Closeness & out\_clo\_wei & Holistic & Efficiency of reach via strong flow channels \\
    In-Eigenvector Centrality & in\_ev & Holistic & Importance based on connections to significant destinations \\
    Out-Eigenvector Centrality & out\_ev & Holistic & Influence through outflows to significant nodes \\
    \hline
    \end{tabular}}
\end{table*}

\begin{table}[!htbp]
\centering
\caption{Sample of Network Measures}
\label{tab:network_measures}
\resizebox{\columnwidth}{!}{
\begin{tabular}{l l r r r r r}
\hline
\textbf{Town} & \textbf{Week} & \textbf{Betweenness} & \textbf{Closeness} & \textbf{Clustering} & \textbf{In-PageRank} & \textbf{Out-PageRank} \\
\hline
Abington & 2020-04-12 & 0.000693 & 0.577778 & 0.663822 & 0.003300 & 0.003119 \\
Abington & 2020-04-19 & 0.000756 & 0.570384 & 0.644578 & 0.003257 & 0.003046 \\
Abington & 2020-04-26 & 0.000739 & 0.576710 & 0.677751 & 0.003388 & 0.003009 \\
Abington & 2020-05-03 & 0.001021 & 0.585366 & 0.646307 & 0.003238 & 0.003279 \\
Abington & 2020-05-10 & 0.000881 & 0.587571 & 0.673452 & 0.003354 & 0.003147 \\
\hline
\hline \\[-1.8ex] 
\end{tabular}}
\end{table}

\begin{table*}[!htbp]
\centering 
\caption{Holistic Network Measures Model Comparison} 
\label{tab:simple_ols_holistic_results} 
\resizebox{\textwidth}{!}{
\begin{tabular}{@{\extracolsep{5pt}}lcccccccccccc} 
\\[-1.8ex]\hline 
\hline \\[-1.8ex] 
\\[-1.8ex] & \multicolumn{12}{c}{} \\ 
& Basic & In-EV & Out-EV & In-PR & Out-PR & In-Clo & Out-Clo & In-Clo-Wei & Out-Clo-Wei & \textbf{Bet} & Bet-Wei & Clustering \\ 
\\[-1.8ex] & (1) & (2) & (3) & (4) & (5) & (6) & (7) & (8) & (9) & (10) & (11) & (12)\\ 
\hline \\[-1.8ex] 
 $MA_t$ & 0.003$^{***}$ & 0.003$^{***}$ & 0.003$^{***}$ & 0.003$^{***}$ & 0.003$^{***}$ & 0.003$^{***}$ & 0.003$^{***}$ & 0.003$^{***}$ & 0.003$^{***}$ & 0.003$^{***}$ & 0.003$^{***}$ & 0.003$^{***}$ \\ 
  & (0.0001) & (0.0001) & (0.0001) & (0.0001) & (0.0001) & (0.0001) & (0.0001) & (0.0001) & (0.0001) & (0.0001) & (0.0001) & (0.0001) \\ 
  & & & & & & & & & & & & \\ 
 in\_ev &  & $-$521.124$^{***}$ &  &  &  &  &  &  &  &  &  &  \\ 
  &  & (110.059) &  &  &  &  &  &  &  &  &  &  \\ 
  & & & & & & & & & & & & \\ 
 out\_ev &  &  & $-$610.422$^{***}$ &  &  &  &  &  &  &  &  &  \\ 
  &  &  & (74.723) &  &  &  &  &  &  &  &  &  \\ 
  & & & & & & & & & & & & \\ 
 in\_pr &  &  &  & 11,787.130$^{***}$ &  &  &  &  &  &  &  &  \\ 
  &  &  &  & (1,039.438) &  &  &  &  &  &  &  &  \\ 
  & & & & & & & & & & & & \\ 
 out\_pr &  &  &  &  & $-$16,643.850$^{***}$ &  &  &  &  &  &  &  \\ 
  &  &  &  &  & (1,913.138) &  &  &  &  &  &  &  \\ 
  & & & & & & & & & & & & \\ 
 in\_clo &  &  &  &  &  & $-$11.902 &  &  &  &  &  &  \\ 
  &  &  &  &  &  & (29.836) &  &  &  &  &  &  \\ 
  & & & & & & & & & & & & \\ 
 out\_clo &  &  &  &  &  &  & $-$239.102$^{***}$ &  &  &  &  &  \\ 
  &  &  &  &  &  &  & (35.153) &  &  &  &  &  \\ 
  & & & & & & & & & & & & \\ 
 in\_clo\_wei &  &  &  &  &  &  &  & $-$0.804$^{***}$ &  &  &  &  \\ 
  &  &  &  &  &  &  &  & (0.079) &  &  &  &  \\ 
  & & & & & & & & & & & & \\ 
 out\_clo\_wei &  &  &  &  &  &  &  &  & $-$1.128$^{***}$ &  &  &  \\ 
  &  &  &  &  &  &  &  &  & (0.115) &  &  &  \\ 
  & & & & & & & & & & & & \\ 
 bet &  &  &  &  &  &  &  &  &  & 26,361.040$^{***}$ &  &  \\ 
  &  &  &  &  &  &  &  &  &  & (639.764) &  &  \\ 
  & & & & & & & & & & & & \\ 
 bet\_wei &  &  &  &  &  &  &  &  &  &  & $-$76.629 &  \\ 
  &  &  &  &  &  &  &  &  &  &  & (91.090) &  \\ 
  & & & & & & & & & & & & \\ 
 clustering &  &  &  &  &  &  &  &  &  &  &  & $-$187.231$^{***}$ \\ 
  &  &  &  &  &  &  &  &  &  &  &  & (21.602) \\ 
  & & & & & & & & & & & & \\ 
 Constant & $-$2.950 & 10.877 & 28.430$^{**}$ & $-$36.375$^{***}$ & 52.386$^{***}$ & 4.205 & 145.974$^{***}$ & 55.599$^{***}$ & 48.252$^{***}$ & $-$27.195$^{**}$ & $-$2.655 & 123.220$^{***}$ \\ 
  & (11.661) & (12.011) & (12.249) & (11.972) & (13.253) & (21.393) & (24.797) & (12.964) & (12.735) & (10.981) & (11.666) & (18.631) \\ 
  & & & & & & & & & & & & \\ 
\hline \\[-1.8ex] 
Predict-$R^2$ & 0.5977 & 0.5983 & 0.5949 & 0.6048 & 0.5977 & 0.5977 & 0.5984 & 0.6009 & 0.6006 & \textbf{0.6616} & 0.5979 & 0.6006 \\ 
Observations & 13,271 & 13,271 & 13,271 & 13,271 & 13,271 & 13,271 & 13,271 & 13,271 & 13,271 & 13,271 & 13,271 & 13,271 \\ 
R$^{2}$ & 0.589 & 0.589 & 0.591 & 0.593 & 0.591 & 0.589 & 0.590 & 0.592 & 0.592 & 0.636 & 0.589 & 0.591 \\ 
Adjusted R$^{2}$ & 0.579 & 0.579 & 0.581 & 0.583 & 0.581 & 0.579 & 0.580 & 0.582 & 0.582 & 0.627 & 0.579 & 0.581 \\ 
\hline 
\hline \\[-1.8ex] 
\multicolumn{12}{l}{All Models include Town Fixed Effect} \\
\multicolumn{12}{l}{$^{*}$p$<$0.1; $^{**}$p$<$0.05; $^{***}$p$<$0.01} \\
\end{tabular}}
\end{table*}

\begin{table*}[!htbp] \centering 
  \caption{Local Network Measures Model Comparison} 
  \label{tab:simple_ols_local_results} 
  \resizebox{\textwidth}{!}{
\begin{tabular}{@{\extracolsep{5pt}}lccccccccc} 
\\[-1.8ex]\hline 
\hline \\[-1.8ex] 
\\[-1.8ex] & \multicolumn{9}{c}{} \\ 
 & Basic & Inweightt & Outweightt & \textbf{Intraweightt} & Intraweightb & LMRt & RLMRt & LMRb & RLMRb \\ 
\\[-1.8ex] & (1) & (2) & (3) & (4) & (5) & (6) & (7) & (8) & (9)\\ 
\hline \\[-1.8ex] 
 MA & 0.003$^{***}$ & 0.003$^{***}$ & 0.002$^{***}$ & 0.002$^{***}$ & 0.003$^{***}$ & 0.003$^{***}$ & 0.003$^{***}$ & 0.003$^{***}$ & 0.003$^{***}$ \\ 
  & (0.0001) & (0.0001) & (0.0001) & (0.0001) & (0.0001) & (0.0001) & (0.0001) & (0.0001) & (0.0001) \\ 
  & & & & & & & & & \\ 
 inweightt &  & $-$0.025$^{***}$ &  &  &  &  &  &  &  \\ 
  &  & (0.001) &  &  &  &  &  &  &  \\ 
  & & & & & & & & & \\ 
 outweightt &  &  & $-$0.047$^{***}$ &  &  &  &  &  &  \\ 
  &  &  & (0.001) &  &  &  &  &  &  \\ 
  & & & & & & & & & \\ 
 intraweightt &  &  &  & $-$0.030$^{***}$ &  &  &  &  &  \\ 
  &  &  &  & (0.0005) &  &  &  &  &  \\ 
  & & & & & & & & & \\ 
 intraweightb &  &  &  &  & $-$0.041$^{***}$ &  &  &  &  \\ 
  &  &  &  &  & (0.001) &  &  &  &  \\ 
  & & & & & & & & & \\ 
 LMRt &  &  &  &  &  & $-$42.908$^{**}$ &  &  &  \\ 
  &  &  &  &  &  & (18.416) &  &  &  \\ 
  & & & & & & & & & \\ 
 RLMRt &  &  &  &  &  &  & $-$62.265$^{***}$ &  &  \\ 
  &  &  &  &  &  &  & (16.663) &  &  \\ 
  & & & & & & & & & \\ 
 LMRb &  &  &  &  &  &  &  & $-$42.443$^{***}$ &  \\ 
  &  &  &  &  &  &  &  & (15.339) &  \\ 
  & & & & & & & & & \\ 
 RLMRb &  &  &  &  &  &  &  &  & $-$15.981 \\ 
  &  &  &  &  &  &  &  &  & (13.784) \\ 
  & & & & & & & & & \\ 
 Constant & $-$2.950 & 77.679$^{***}$ & 160.073$^{***}$ & 151.159$^{***}$ & 148.636$^{***}$ & 22.572 & 57.281$^{***}$ & 15.596 & 11.754 \\ 
  & (11.661) & (11.517) & (11.593) & (10.597) & (11.177) & (15.997) & (19.891) & (13.447) & (17.228) \\ 
  & & & & & & & & & \\ 
\hline \\[-1.8ex] 
Predict-$R^2$ & 0.5977 & 0.6577 & 0.6428 & \textbf{0.7055} & 0.6507 & 0.5979 & 0.5987 & 0.5978 & 0.5978 \\ 
Observations & 13,271 & 13,271 & 13,271 & 13,271 & 13,271 & 13,271 & 13,271 & 13,271 & 13,271 \\ 
R$^{2}$ & 0.589 & 0.618 & 0.638 & 0.680 & 0.652 & 0.589 & 0.589 & 0.589 & 0.589 \\ 
Adjusted R$^{2}$ & 0.579 & 0.609 & 0.630 & 0.672 & 0.643 & 0.579 & 0.579 & 0.579 & 0.579 \\ 
\hline 
\hline \\[-1.8ex] 
\multicolumn{9}{l}{All Models include Town Fixed Effect} \\
\multicolumn{9}{l}{$^{*}$p$<$0.1; $^{**}$p$<$0.05; $^{***}$p$<$0.01} \\
\end{tabular}} 
\end{table*}

When considering \textbf{holistic network measures} (Table \ref{tab:simple_ols_holistic_results}), we found that \textbf{Betweenness Centrality} (Bet) provided the most substantial improvement in predictive power. Including this metric increased the predictive $R^2$ to 66.16\%, demonstrating its strong ability to capture key aspects of disease spread across the network.

Among the \textbf{local network measures} (Table \ref{tab:simple_ols_local_results}), \textbf{Intra-Town Flow} (Intraweightt) emerged as the top performer. This metric, which quantifies the internal mobility (flow of people) within each town (i.e., movement between census block groups within the same town), significantly boosted the predictive $R^2$ to 70.55\%. This highlights the critical role of localized mobility patterns in forecasting COVID-19 dynamics.


\newpage
\subsection{Spatial Regression Analysis}

\begin{figure*}[!htbp]
    \centering
    \includegraphics[width=0.9\textwidth]{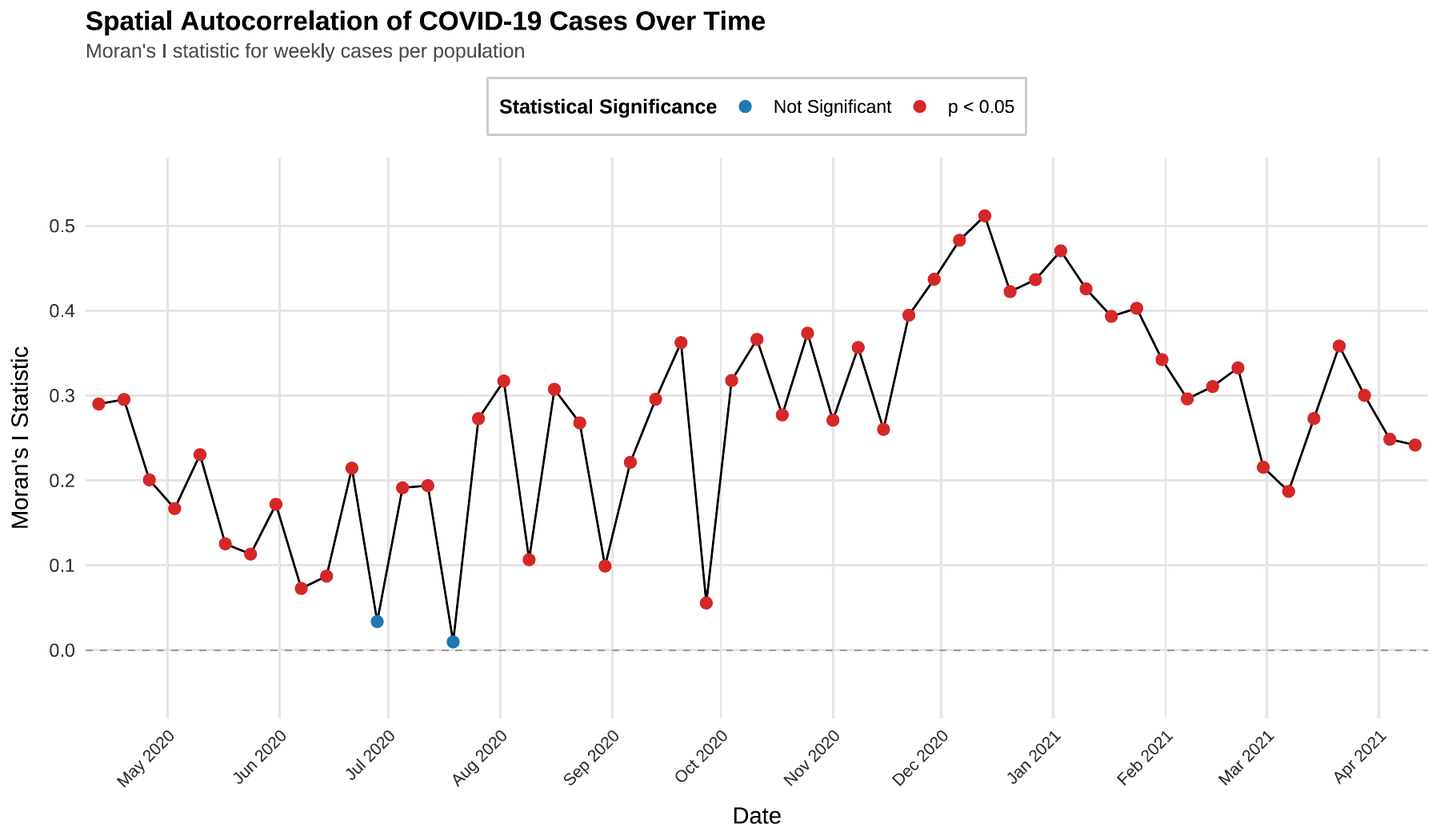}
    \caption{Temporal evolution of spatial autocorrelation in COVID transmission across Massachusetts towns. The Moran's I statistic is calculated weekly for population-normalized case rates from April 2020 to April 2021. Red points indicate statistically significant spatial clustering (p < 0.05), while blue points represent non-significant values. The dashed horizontal line at zero represents the null hypothesis of no spatial autocorrelation. Higher Moran's I values indicate stronger spatial clustering, with neighboring towns exhibiting similar case rates per population. }

    \label{fig:moran_test}
\end{figure*}

To implement spatial analysis, we initially performed Moran's I test as the classic test for measuring spatial autocorrelation \cite{Moran_1950}. Moran's I test is designed for fixed networks and does not capture temporal dependencies. Thus, we performed this test for each week, and visualized its evolution over time in Figure \ref{fig:moran_test}. We normalized case counts by town population (using census data) to account for population differences. As expected, Moran's I values for absolute case counts are much lower due to the adjacency of many high and low-populated towns. However, once the population is accounted for, significant spatial correlation emerges.

The temporal evolution of Moran's I values reveals important patterns in the spatial distribution of COVID-19 cases across Massachusetts towns throughout 2020-2021. In the early months of the pandemic (April-May 2020), moderate spatial clustering was already present, indicating that even initially, the virus wasn't uniformly distributed across all towns. As we moved into Summer 2020, clustering decreased slightly, with some periods showing non-significant spatial patterns, suggesting a more random distribution of cases across communities. This may be due to restrictive lockdowns. However, by late 2020 through January 2021, Moran's I values reached their peak (above 0.5), demonstrating that significant geographic hotspots had formed - some regions were experiencing substantially higher case rates per population while others remained relatively protected.
This increased spatial clustering during the winter surge highlights how transmission patterns evolved to create distinct regional risk patterns across Massachusetts. By spring 2021, the Moran's I values declined toward moderate levels again, potentially reflecting the impact of targeted vaccination efforts, changes in mobility patterns, or a more even distribution of cases as previously less-affected areas caught up. The consistently significant spatial autocorrelation throughout most of the study period underscores that COVID-19 transmission wasn't random but followed geographic patterns that intensified during surge periods and relaxed during periods of lower overall transmission.

Spatial Regression results are shown in Table \ref{tab:spatial_models}.
The progression of models reveals important dynamics in COVID-19 transmission patterns across Massachusetts towns. Model 1 demonstrates that betweenness centrality alone has a strong positive association with next week's cases, suggesting that a town's position in the mobility network significantly influences disease spread. Model 2 shows that the previous week's total cases across all towns also positively predicted local transmission. When combined in Model 3, both predictors maintain their significance and direction, indicating they capture different aspects of transmission dynamics.

Model 4, which introduces the interaction term, provides the most nuanced understanding of how network position and existing case load worked together to drive the pandemic. The negative base coefficient for betweenness centrality, coupled with the strongly positive interaction term, reveals a crucial pattern: towns with high centrality only became amplifiers of disease transmission when existing case loads were substantial. This explains the apparent contradiction in the visualization showing Boston (high centrality) with high case rates - at high case levels, the positive interaction effect dominated. The significant spatial parameters across all models (positive $\rho$ and negative $\lambda$) indicate that unobserved factors exhibited spatial clustering, while the negative spatial lag suggests some competitive effects between neighboring towns, possibly due to reduced travel between affected areas or localized containment measures. The narrowing residual distribution in Model 4 further supports its superior explanatory power in capturing these complex spatial-temporal interactions in COVID-19 transmission.

\begin{table}[!htbp]
\centering
\caption{Spatial Panel Fixed Effects SARAR Models for COVID-19 Cases}
\label{tab:spatial_models}
\resizebox{0.7\textwidth}{!}{
\begin{tabular}{lcccc}
\hline\hline
 & \multicolumn{4}{c}{Dependent Variable: Next Week COVID-19 Cases (y)} \\
 & Model 1 & Model 2 & Model 3 & Model 4 \\
\hline
bet & 24780.54*** & & 26264.00*** & -6437.30*** \\
 & (514.67) & & (538.85) & (459.51) \\
 & & & & \\
MA & & 0.00407*** & 0.00345*** & 0.00041*** \\
 & & (0.00016) & (0.00014) & (0.00010) \\
 & & & & \\
bet:MA & & & & 1.374*** \\
 & & & & (0.0114) \\
 & & & & \\
\hline
Spatial Parameters & & & & \\
\hline
Spatial error ($\rho$) & 0.747*** & 0.605*** & 0.543*** & 0.575*** \\
 & (0.009) & (0.017) & (0.020) & (0.013) \\
 & & & & \\
Spatial lag ($\lambda$) & -0.484*** & -0.370*** & -0.240*** & -0.167*** \\
 & (0.022) & (0.031) & (0.031) & (0.017) \\
\hline
\multicolumn{5}{l}{Residual Summary} \\
\hline
Min & -959.45 & -1071.16 & -906.91 & -676.01 \\
1st Quartile & -23.87 & -18.69 & -17.78 & -9.27 \\
Median & -4.99 & 1.14 & 1.75 & 0.10 \\
3rd Quartile & 10.95 & 20.08 & 18.61 & 8.31 \\
Max & 2772.73 & 2734.27 & 2616.70 & 1574.06 \\
\hline\hline
\multicolumn{4}{l}{\footnotesize{$* p<0.1, ** p<0.05, *** p<0.01$. Standard errors in parentheses.}} \\
\multicolumn{4}{l}{\footnotesize{Note: All models include town-specific fixed effects.}} \\
\end{tabular}}
\end{table}

\newpage
\section{Agent-Based Modeling (ABM)}

Agent-based models (ABMs) have become a central tool in studying infectious
disease dynamics and human behavior, offering a way to capture heterogeneous
interactions under controlled conditions. Several recent studies have employed
ABMs to understand COVID-19 spread in urban contexts, to combine mobility traces
with epidemic simulations, and to evaluate public-health interventions
\cite{pesavento_abm, jalayer2020cov, silva2020covid, hunter2020hybrid, miksch2019should, muller2021predicting}.
Building on this literature, our ABM serves as a mechanistic triangulation tool
rather than a minor validation step. The reduced-form regressions rely on
imperfect case reports and aggregated mobility metrics; the ABM, by contrast,
offers full observability of infections, movements, and network structure under
known transmission rules. It lets us test whether the conditional forecasting
patterns observed in the Massachusetts data, namely large gains from
macro--network interactions when local histories are absent and minimal added
value once autoregression is available, arise from a general transmission
mechanism rather than from specification artifacts.

The defining feature of our design is that the ABM is \emph{deliberately
uncalibrated} to the Massachusetts data. Rather than reproducing the observed
mobility system, it simulates epidemics on \emph{generic} synthetic networks
with known structure. If the regime-dependent pattern were specific to the
Massachusetts setting, generic simulations would fail to reproduce it; the lack
of calibration is therefore what gives the exercise its inferential force. The
implementation is the \texttt{absim} Python package; we summarize the model
below and give exact parameters in Table~\ref{tab:abm_params}.

\subsection{Synthetic Network Families}
Each town is a node of a directed, weighted network; edge weight $w_{ij}$ is the
flow from town $i$ to town $j$ and the self-loop $w_{ii}$ is within-town mixing.
To test topological generality we generate networks from five families at fixed
$N$ and mean degree: Barab\'asi--Albert (scale-free; baseline), Erd\H{o}s--R\'enyi,
Watts--Strogatz (small-world), stochastic-block (modular), and spatial
random-geometric. All families share one weighting scheme: node ``sizes'' are
drawn lognormal with spread set by a \emph{heterogeneity knob}, and edge weights
follow a gravity rule $w_{ij}\propto (\text{size}_i\,\text{size}_j)/d_{ij}^{\,\rho}$
(with Euclidean distance $d_{ij}$ for the spatial family and $d_{ij}\equiv1$
otherwise); self-loops are a fixed fraction of each node's throughput. For every
generated network we record topological-heterogeneity summaries (betweenness
variance, degree CV, modularity) used in the H2 analysis.

\subsection{Epidemic Dynamics}
We use a metapopulation stochastic SIR(S) process rather than literal agent
movement, which is far cheaper for the thousands of replicate runs the sweeps
require. Each town holds a closed sub-population $(S_i,I_i,R_i)$. Per time step
$\mathrm{d}t$ (daily), susceptibles in town $i$ are exposed to a
\emph{mobility-weighted average} of neighbour prevalence. Crucially, we separate
the mixing \emph{pattern} from the contact \emph{intensity}:
\begin{equation}
\mathrm{foi}_i \;=\; \beta\,\underbrace{c_i}_{\text{intensity}}\sum_j \underbrace{\widehat{W}_{ij}}_{\text{pattern}}\,\frac{I_j}{P_j},
\qquad
\widehat{W}=\text{rownorm}(W_t),\quad c_i=\frac{\textstyle\sum_j W_{t,ij}}{\textstyle\sum_j W_{0,ij}} .
\label{eq:abm_foi}
\end{equation}
New infections and recoveries are binomial draws,
$\mathrm{Bin}(S_i,1-e^{-\mathrm{foi}_i\,\mathrm{d}t})$ and
$\mathrm{Bin}(I_i,1-e^{-\gamma\,\mathrm{d}t})$; with waning rate $\omega>0$,
recovered individuals return to $S$ at rate $\omega$ (SIRS), which sustains
multi-wave dynamics across the window. The intensity term $c_i$ is essential:
because $W_t=W_0$ scaled per row by the mobility multiplier, a uniform row
rescaling would cancel under row-normalization, so without $c_i$ the mobility
level would have no dynamical effect. Daily new infections are aggregated to
weeks to give $y_{i,t}$, and $MA_t=\sum_i y_{i,t}$.

\subsection{Endogenous versus Exogenous Mobility}
In the \emph{endogenous} mode the weekly network responds to prevalence through a
Lipschitz map $M_{t+1}=\Phi(y_t,M_t)$: between-town flows are suppressed more
strongly than within-town activity (responsiveness $\kappa>\kappa_{\text{self}}$),
so high prevalence reshapes the topology, not merely its overall level. The
\emph{exogenous} mode applies a prevalence-independent schedule constructed as a
circular time-shift of the endogenous multiplier trajectory; this matches the
endogenous marginal variance \emph{and} temporal smoothness while decoupling
mobility from the contemporaneous epidemic, isolating the role of the prevalence
coupling (used in the operator-drift test below).

\subsection{Network Features and Forecasting}
Per week and town we extract flow betweenness (shortest-path betweenness on the
directed graph with distance $1/w$), intra-town flow (self-loop weight), and
in-strength $r_{j,t}=\sum_i w_{ij,t}$. We then run the \emph{same} forecasting
battery as the empirical analysis (Panels~A and~B of the main-text tables), with
town fixed effects estimated by the within transformation, feature
standardization fit on training weeks only, and blocked time-wise
cross-validation (contiguous week blocks held out in turn). The primary metric
is Predict-$R^2$ on held-out observations; predicted weekly case counts are
clipped to be non-negative and bounded by a multiple of the training maximum,
which prevents linear extrapolation from blowing up when a held-out block
contains an epidemic peak. We make no attempt to calibrate ABM parameters to the
empirical coefficients or $R^2$ values: agreement, where it occurs, reflects
shared structure rather than tuning.

\begin{table}[t]
\centering
\caption{Baseline ABM parameters (\texttt{absim} fast/full configurations).}
\label{tab:abm_params}
\small
\begin{tabular}{@{}lll@{}}
\toprule
Symbol & Meaning & Baseline value \\
\midrule
$N$ & number of towns & 60 (dev) / 40 (production) \\
$\bar k$ & mean degree & 6 \\
$h$ & heterogeneity knob (lognormal $\sigma$) & 1.0 \\
$f_{\text{self}}$ & self-loop fraction of throughput & 0.3 \\
$\beta$ & transmissibility ($R_0\!\approx\!\beta/\gamma$) & 0.18 \\
$\gamma$ & recovery rate (per day) & 0.10 \\
$\omega$ & waning rate (per day) & 0.02 \\
$\kappa,\ \kappa_{\text{self}}$ & mobility responsiveness (between, within) & 5,\ 1 \\
$\alpha$ & local vs.\ global prevalence weight & 0.8 \\
$T$ & weeks simulated & 40 \\
$P$ & mean town population & 10{,}000 \\
-- & CV blocks / burn-in weeks / horizons & 5 / 4 / $\{1,2,3,4\}$ \\
-- & stochastic replicates (seeds) & 30 (production) \\
\bottomrule
\end{tabular}
\end{table}

\subsection{Experiment Designs (H1--H5)}
Using this model we evaluate four predictions; results and figures are in the
main text (Generalizability section). Each cell is averaged over 30 random seeds
and reported as mean $\pm$ 95\% confidence interval.

\begin{itemize}
\item \textbf{H1 (generality across families).} Run the battery on each of the
five network families and compare the data-limited gain (best macro--network
model minus the macro-only baseline) with the data-rich gain (best AR+network
model minus the AR baseline).
\item \textbf{H2 (heterogeneity).} Within the Barab\'asi--Albert family, sweep the
heterogeneity knob and relate the data-limited gain to the measured betweenness
variance of the generated networks.
\item \textbf{H3 (operator drift; Proposition on multi-step behavior).} Contrast
endogenous mobility against the variance- and smoothness-matched exogenous null,
measuring operator drift $\lVert M_{t+k}-M_t\rVert/\lVert M_t\rVert$ and the
scale-fitted error of the iterated operator predictor. A rapid-mobility-shift
regime (a policy-like threshold lockdown response) is used to enlarge drift.
\item \textbf{H5 (channel degradation).} Degrade the autoregressive channel three
ways (delay $y_{i,t-k}$; additive noise across an SNR grid; missingness for a
fraction of towns) and track the substitution value, the network lift over the
AR baseline, as a continuous function of degradation.
\end{itemize}
We also examined an $R_0$ sweep (varying $\beta/\gamma$); we report it only as a
scope note, since the manuscript's ``rapid-change phase'' claim concerns
within-run epidemic phases (operator drift) rather than cross-run transmissibility.

\subsection{Reproducibility}
The model, experiments, and figures are produced by the \texttt{absim} package.
The full run regenerates every reported number and figure from a single command,
\texttt{python -m absim.make\_results configs/full.yaml 30}, with the fast
development configuration (\texttt{configs/fast.yaml}) used for iteration; one
seed parameter threads through all random number generation, and raw per-run case
trajectories are saved so the analysis can be re-run without re-simulating. A
test suite covers feature extraction, the Predict-$R^2$ computation, and the
leakage controls.

\section{Multi-period Analysis}

\subsection{Temporal Heterogeneity in Pandemic Dynamics}

In our analysis of COVID-19 transmission dynamics, we divided the data into several distinct periods to capture different stages of the pandemic. The "Whole Period" encompasses the entire dataset from April 2020 to April 2021, allowing us to understand the overall patterns. For more detailed analysis, we separated two major waves of infection: the "First Wave" (April-June 2020) representing the initial outbreak, and the "Second Wave" (October 2020-April 2021) representing the more prolonged winter surge. We further subdivided the second wave into "Rising" (October 2020-January 2021) and "Falling" (January-April 2021) phases to examine potential differences in transmission dynamics during growth versus decline periods. Additionally, we identified a "Smooth Period" between the two major waves (June-October 2020), characterized by relatively stable case numbers, which serves as a useful comparison to the more volatile wave periods. The "Two Waves" analysis combines data from both wave periods while excluding the smooth period between them.

\begin{table*}[!htbp]
\centering
\caption{Definition of Periods}
\label{tab:periods_definitions}
\resizebox{\textwidth}{!}{
\begin{tabular}{@{}llp{7cm}@{}} 
\toprule 
\textbf{Period Name} & \textbf{Date Range} & \textbf{Description} \\
\midrule 
\addlinespace[0.5em] 
Whole Period & April 13, 2020 - April 12, 2021 & The entire study period, covering all waves and intermediate periods \\
\addlinespace[0.8em] 
Two Waves & \begin{tabular}[t]{@{}l@{}}April 13, 2020 - June 22, 2020 and\\October 19, 2020 - April 12, 2021\end{tabular} & Combines the first and second wave periods, excluding the smooth period between them \\
\addlinespace[0.8em]
Smooth Period & June 22, 2020 - October 19, 2020 & Period between waves characterized by relatively stable case numbers \\
\addlinespace[0.8em]
First Wave & April 13, 2020 - June 22, 2020 & Initial COVID-19 outbreak in Massachusetts \\
\addlinespace[0.8em]
Second Wave & October 19, 2020 - April 12, 2021 & Winter surge of COVID-19, characterized by higher case numbers and longer duration than the first wave \\
\addlinespace[0.8em]
Second Wave Rising & October 19, 2020 - January 4, 2021 & Growth phase of the second wave with increasing case numbers \\
\addlinespace[0.8em]
Second Wave Falling & January 11, 2021 - April 12, 2021 & Decline phase of the second wave as case numbers gradually decreased \\
\addlinespace[0.5em] 
\bottomrule 
\end{tabular}
}
\end{table*}

\begin{table}[!htbp]
\centering
\caption{COVID-19 Prediction Models. All models include town fixed effects to account for town-specific variations. MA is the total cases across all towns in Massachusetts, and intraweightt and bet are the top local and holistic network metrics, respectively. The target variable, \textit{y}, is also next week's COVID-19 cases in each town}
\label{tab:model_equations}
\resizebox{0.6\textwidth}{!}{
\begin{tabular}{ll}
\hline
\textbf{Model Name} & \textbf{Equation} \\
\hline
Basic & $y \sim \text{MA}$ \\
Mobility & $y \sim \text{MA} + \text{intraweightt}$ \\
Mobility+Int & $y \sim \text{MA} + \text{intraweightt} + \text{intraweightt} \times \text{MA}$ \\
Network & $y \sim \text{MA} + \text{bet}$ \\
Network+Int & $y \sim \text{MA} + \text{bet} + \text{bet} \times \text{MA}$ \\
Combined & $y \sim \text{MA} + \text{intraweightt} + \text{bet}$ \\
Combined+Int & $y \sim \text{MA} + \text{intraweightt} + \text{intraweightt} \times \text{MA} + \text{bet} + \text{bet} \times \text{MA}$ \\
\hline
\end{tabular}}
\end{table}

Table \ref{tab:periods_definitions} provides a clear definition of each period analyzed in the study, including exact date ranges and descriptive explanations of what each period represents in pandemic dynamics. Table \ref{tab:model_equations} presents our model specifications for predicting next week's COVID-19 cases in each town. All models include town fixed effects to account for location-specific variations. The table shows the equation for each regression model, using the top-performing local and holistic network centrality metrics, which are in-town movements and betweenness centrality, respectively. Models with interaction terms (denoted by "+Int") include multiplicative interactions between MA and the respective network measures, allowing the effect of these measures to vary with the current infection level. The combined models integrate both local and holistic network metrics to evaluate their joint predictive power.

\begin{table*}[!htbp] \centering 
  \caption{COVID-19 Transmission Models: Comparison Across the Whole Period and Wave Periods. This table presents regression results comparing models estimated on data from the entire study period versus only the wave periods. All models include town fixed effects. The mobility and network models both substantially improve model fit compared to the basic model, which only includes state-level cases (MA). Standard errors in parentheses.} 
  \label{tab:whole_vs_waves} 
\resizebox{\textwidth}{!}{
\begin{tabular}{@{\extracolsep{5pt}}lcccccccc} 
\hline
               & \begin{tabular}{c} Whole Period \\ Basic Model \end{tabular} & \begin{tabular}{c} Whole Period \\ Mobility Model \end{tabular} & \begin{tabular}{c} Whole Period \\ Network Model \end{tabular} & \begin{tabular}{c} Whole Period \\ Combined Model \end{tabular} & \begin{tabular}{c} Two Waves \\ Basic Model \end{tabular} & \begin{tabular}{c} Two Waves \\ Mobility Model \end{tabular} & \begin{tabular}{c} Two Waves \\ Network Model \end{tabular} & \begin{tabular}{c} Two Waves \\ Combined Model \end{tabular}  \\
\hline
 MA & 0.003$^{***}$ & $-$0.001$^{***}$ & $-$0.0002$^{***}$ & $-$0.001$^{***}$ & 0.003$^{***}$ & $-$0.0004$^{***}$ & $-$0.00002 & $-$0.0003$^{***}$ \\ 
  & (0.0001) & (0.00004) & (0.0001) & (0.00004) & (0.0001) & (0.0001) & (0.0001) & (0.0001) \\ 
  & & & & & & & & \\ 
 intraweightt &  & $-$0.009$^{***}$ &  & $-$0.010$^{***}$ &  & $-$0.014$^{***}$ &  & $-$0.017$^{***}$ \\ 
  &  & (0.0003) &  & (0.0004) &  & (0.001) &  & (0.001) \\ 
  & & & & & & & & \\ 
 bet &  &  & $-$5,396.817$^{***}$ & 1,798.626$^{***}$ &  &  & $-$13,774.390$^{***}$ & 753.507 \\ 
  &  &  & (577.593) & (458.042) &  &  & (817.754) & (669.259) \\ 
  & & & & & & & & \\ 
 MA:intraweightt &  & 0.00000$^{***}$ &  & 0.00000$^{***}$ &  & 0.00000$^{***}$ &  & 0.00000$^{***}$ \\ 
  &  & (0.000) &  & (0.000) &  & (0.000) &  & (0.000) \\ 
  & & & & & & & & \\ 
 MA:bet &  &  & 1.409$^{***}$ & $-$0.299$^{***}$ &  &  & 1.312$^{***}$ & $-$0.491$^{***}$ \\ 
  &  &  & (0.014) & (0.021) &  &  & (0.016) & (0.028) \\ 
  & & & & & & & & \\ 
 Constant & $-$2.950 & 41.107$^{***}$ & 17.879$^{**}$ & 41.225$^{***}$ & $-$6.499 & 66.155$^{***}$ & 33.687$^{***}$ & 68.218$^{***}$ \\ 
  & (11.661) & (6.587) & (8.277) & (6.589) & (15.158) & (9.454) & (11.340) & (9.287) \\ 
  & & & & & & & & \\ 
\hline \\[-1.8ex] 
Predict-$R^2$ & 0.5977 & 0.8909 & 0.8290 & 0.8924 & 0.7502 & 0.9171 & 0.8784 & 0.9170 \\ 
Observations & 13,271 & 13,271 & 13,271 & 13,271 & 8,764 & 8,764 & 8,764 & 8,764 \\ 
R$^{2}$ & 0.589 & 0.878 & 0.794 & 0.880 & 0.661 & 0.880 & 0.812 & 0.885 \\ 
Adjusted R$^{2}$ & 0.579 & 0.875 & 0.789 & 0.877 & 0.648 & 0.876 & 0.805 & 0.881 \\ 
\hline 
\hline \\[-1.8ex] 
\textit{Note:}  & \multicolumn{8}{l}{$^{*}$p$<$0.1; $^{**}$p$<$0.05; $^{***}$p$<$0.01} \\ 
\end{tabular}}
\end{table*}

Table \ref{tab:whole_vs_waves} presents regression results comparing models estimated on data from the entire study period versus only the wave periods. The mobility and network models both substantially improve model fit compared to the basic model, which only includes state-level cases (MA).

\begin{table*}[!htbp] \centering 
  \caption{COVID-19 Transmission Dynamics: Wave Periods vs. Smooth Period Comparison} 
  \label{tab:wave_vs_smooth} 
\resizebox{\textwidth}{!}{
\begin{tabular}{@{\extracolsep{5pt}}lcccccccc} 
\hline
               & \begin{tabular}{c} Two Waves \\ Basic Model \end{tabular} & \begin{tabular}{c} Two Waves \\ Mobility Model \end{tabular} & \begin{tabular}{c} Two Waves \\ Network Model \end{tabular} & \begin{tabular}{c} Two Waves \\ Combined Model \end{tabular} & \begin{tabular}{c} Smooth Period \\ Basic Model \end{tabular} & \begin{tabular}{c} Smooth Period \\ Mobility Model \end{tabular} & \begin{tabular}{c} Smooth Period \\ Network Model \end{tabular} & \begin{tabular}{c} Smooth Period \\ Combined Model \end{tabular}  \\
\hline
 MA & 0.003$^{***}$ & $-$0.0004$^{***}$ & $-$0.00002 & $-$0.0003$^{***}$ & 0.004$^{***}$ & $-$0.001$^{***}$ & $-$0.002$^{***}$ & $-$0.001$^{***}$ \\ 
  & (0.0001) & (0.0001) & (0.0001) & (0.0001) & (0.0002) & (0.0001) & (0.0002) & (0.0002) \\ 
  & & & & & & & & \\ 
 intraweightt &  & $-$0.014$^{***}$ &  & $-$0.017$^{***}$ &  & $-$0.002$^{***}$ &  & $-$0.003$^{***}$ \\ 
  &  & (0.001) &  & (0.001) &  & (0.0002) &  & (0.0002) \\ 
  & & & & & & & & \\ 
 bet &  &  & $-$13,774.390$^{***}$ & 753.507 &  &  & $-$3,013.546$^{***}$ & 1,625.372$^{***}$ \\ 
  &  &  & (817.754) & (669.259) &  &  & (483.155) & (400.665) \\ 
  & & & & & & & & \\ 
 MA:intraweightt &  & 0.00000$^{***}$ &  & 0.00000$^{***}$ &  & 0.00000$^{***}$ &  & 0.00000$^{***}$ \\ 
  &  & (0.000) &  & (0.000) &  & (0.000) &  & (0.00000) \\ 
  & & & & & & & & \\ 
 MA:bet &  &  & 1.312$^{***}$ & $-$0.491$^{***}$ &  &  & 2.618$^{***}$ & $-$0.945$^{***}$ \\ 
  &  &  & (0.016) & (0.028) &  &  & (0.052) & (0.087) \\ 
  & & & & & & & & \\ 
 Constant & $-$6.499 & 66.155$^{***}$ & 33.687$^{***}$ & 68.218$^{***}$ & $-$3.614 & 8.845$^{**}$ & 7.370$^{*}$ & 8.702$^{**}$ \\ 
  & (15.158) & (9.454) & (11.340) & (9.287) & (5.493) & (3.599) & (4.256) & (3.563) \\ 
  & & & & & & & & \\ 
\hline \\[-1.8ex] 
Predict-$R^2$ & 0.7502 & 0.9171 & 0.8784 & 0.9170 & 0.7299 & 0.8724 & 0.8012 & 0.8806 \\ 
Observations & 8,764 & 8,764 & 8,764 & 8,764 & 4,256 & 4,256 & 4,256 & 4,256 \\ 
R$^{2}$ & 0.661 & 0.880 & 0.812 & 0.885 & 0.725 & 0.892 & 0.837 & 0.895 \\ 
Adjusted R$^{2}$ & 0.648 & 0.876 & 0.805 & 0.881 & 0.703 & 0.883 & 0.824 & 0.887 \\ 
\hline 
\hline \\[-1.8ex] 
\textit{Note:}  & \multicolumn{8}{l}{$^{*}$p$<$0.1; $^{**}$p$<$0.05; $^{***}$p$<$0.01} \\ 
\end{tabular}} 
\end{table*} 

Table \ref{tab:wave_vs_smooth} compares model performance across wave periods and the smooth period between waves. The mobility and network models show higher explanatory power compared to the basic model in both periods, with the combined model performing best overall.

\begin{table*}[!htbp] \centering 
  \caption{COVID-19 Transmission Dynamics: First Wave vs. Second Wave Comparison} 
  \label{tab:fir_vs_sec_wave}
\resizebox{\textwidth}{!}{
\begin{tabular}{@{\extracolsep{5pt}}lcccccccc} 
\hline
               & \begin{tabular}{c} First Wave \\ Basic Model \end{tabular} & \begin{tabular}{c} First Wave \\ Mobility Model \end{tabular} & \begin{tabular}{c} First Wave \\ Network Model \end{tabular} & \begin{tabular}{c} First Wave \\ Combined Model \end{tabular} & \begin{tabular}{c} Second Wave \\ Basic Model \end{tabular} & \begin{tabular}{c} Second Wave \\ Mobility Model \end{tabular} & \begin{tabular}{c} Second Wave \\ Network Model \end{tabular} & \begin{tabular}{c} Second Wave \\ Combined Model \end{tabular}  \\
\hline
 MA & 0.002$^{***}$ & $-$0.001$^{***}$ & $-$0.0003$^{**}$ & $-$0.001$^{***}$ & 0.003$^{***}$ & $-$0.001$^{***}$ & $-$0.0001 & $-$0.0005$^{***}$ \\ 
  & (0.0002) & (0.0001) & (0.0001) & (0.0001) & (0.0001) & (0.0001) & (0.0001) & (0.0001) \\ 
  & & & & & & & & \\ 
 intraweightt &  & $-$0.005$^{***}$ &  & $-$0.008$^{***}$ &  & $-$0.008$^{***}$ &  & $-$0.012$^{***}$ \\ 
  &  & (0.001) &  & (0.001) &  & (0.001) &  & (0.001) \\ 
  & & & & & & & & \\ 
 bet &  &  & 1,676.885 & 3,127.852$^{***}$ &  &  & $-$33,178.570$^{***}$ & $-$5,618.066$^{***}$ \\ 
  &  &  & (1,050.249) & (742.732) &  &  & (1,340.306) & (1,375.413) \\ 
  & & & & & & & & \\ 
 MA:intraweightt &  & 0.00000$^{***}$ &  & 0.00000$^{***}$ &  & 0.00000$^{***}$ &  & 0.00000$^{***}$ \\ 
  &  & (0.00000) &  & (0.00000) &  & (0.000) &  & (0.00000) \\ 
  & & & & & & & & \\ 
 MA:bet &  &  & 1.129$^{***}$ & $-$0.331$^{***}$ &  &  & 1.237$^{***}$ & $-$0.335$^{***}$ \\ 
  &  &  & (0.038) & (0.041) &  &  & (0.021) & (0.043) \\ 
  & & & & & & & & \\ 
 Constant & $-$6.688 & 20.786$^{**}$ & 2.434 & 29.902$^{***}$ & 3.308 & 45.819$^{***}$ & 68.518$^{***}$ & 62.090$^{***}$ \\ 
  & (13.246) & (8.669) & (9.193) & (8.623) & (17.962) & (13.626) & (14.347) & (13.551) \\ 
  & & & & & & & & \\ 
\hline \\[-1.8ex] 
Predict-$R^2$ & 0.5328 & 0.8711 & 0.8073 & 0.8663 & 0.8125 & 0.9166 & 0.8884 & 0.9201 \\ 
Observations & 2,504 & 2,504 & 2,504 & 2,504 & 6,260 & 6,260 & 6,260 & 6,260 \\ 
R$^{2}$ & 0.604 & 0.907 & 0.812 & 0.910 & 0.790 & 0.893 & 0.868 & 0.896 \\ 
Adjusted R$^{2}$ & 0.547 & 0.893 & 0.785 & 0.896 & 0.778 & 0.887 & 0.861 & 0.890 \\ 
\hline 
\hline \\[-1.8ex] 
\textit{Note:}  & \multicolumn{8}{l}{$^{*}$p$<$0.1; $^{**}$p$<$0.05; $^{***}$p$<$0.01} \\ 
\end{tabular}} 
\end{table*}

Table \ref{tab:fir_vs_sec_wave} compares transmission dynamics between the first and second COVID-19 waves. The second wave models show higher coefficients for network measures, suggesting increased importance of network centrality during the winter surge. Another reason could be due to our incomplete data for the first wave, as towns started reporting town-level case numbers after the start of the first wave.

\begin{figure*}[!htbp]
    \centering
    \includegraphics[width=\linewidth]{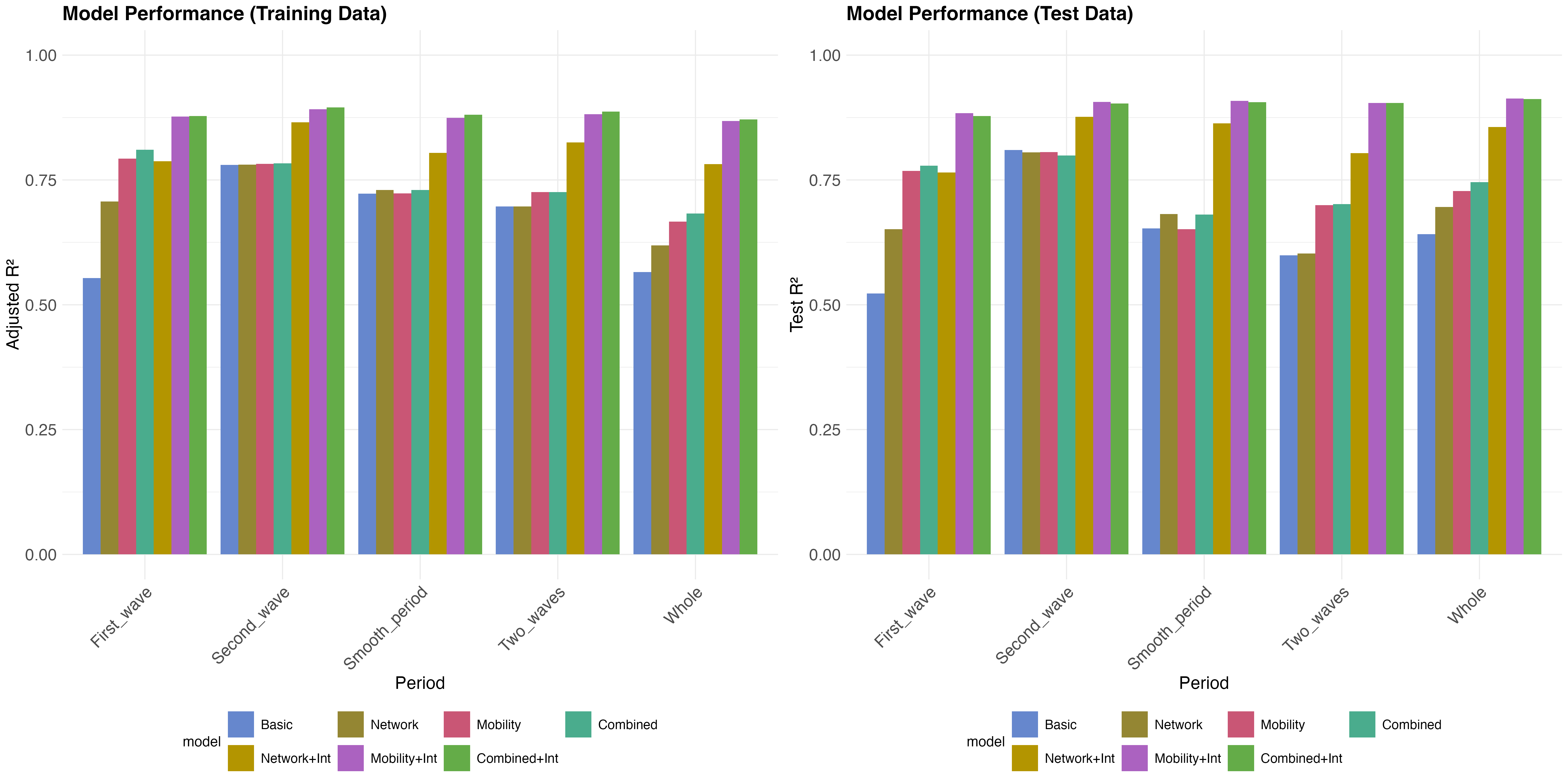}
    \caption{All Model Performances over different periods}
    \label{fig:expanded_model_performance}
\end{figure*}

Figure \ref{fig:expanded_model_performance} demonstrates the consistent superiority of network-enhanced models across multiple pandemic periods. The training and test data results reveal that models incorporating network centrality measures (shown in green) significantly outperform the basic model (gray bars) across all analyzed periods. Notably, the models with interaction terms (darker shades) achieve the highest performance levels, with the 'Combined+Int' model consistently reaching adjusted R² values above 0.85 in most periods. This pattern holds remarkably well across both training and test data, indicating that network-based predictive advantages are robust and not merely artifacts of overfitting.

\begin{figure*}[!htbp]
    \centering
    \includegraphics[width=\linewidth]{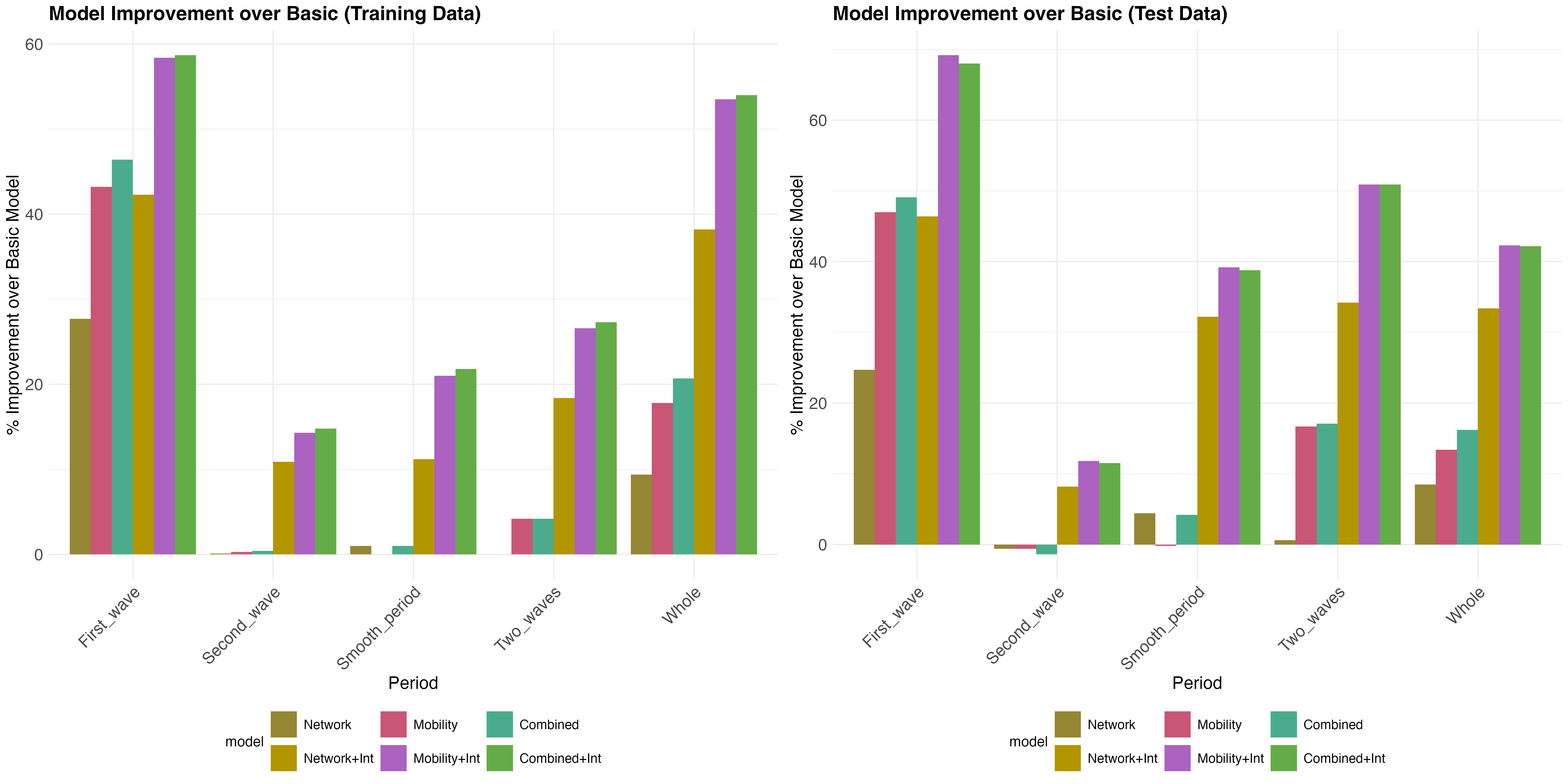}
    \caption{Model Improvements compared to the baseline model without network metrics}
    \label{fig:all_model_improvements}
\end{figure*}

Figure \ref{fig:all_model_improvements} illustrates the comprehensive advantage of network-enhanced models over basic models. Models incorporating network centrality measures show improvements of 20-60\% over the basic model across different periods, with interaction terms providing additional substantial gains. The 'First wave' period shows the most dramatic improvements, suggesting that network effects were particularly pronounced during the initial phase of the pandemic when behavioral and policy responses were still developing. The sustained advantages across both training and test data confirm that network measures capture genuine epidemiological signals rather than spurious correlations.

\begin{figure*}[!htbp]
    \centering
    \includegraphics[width=\linewidth]{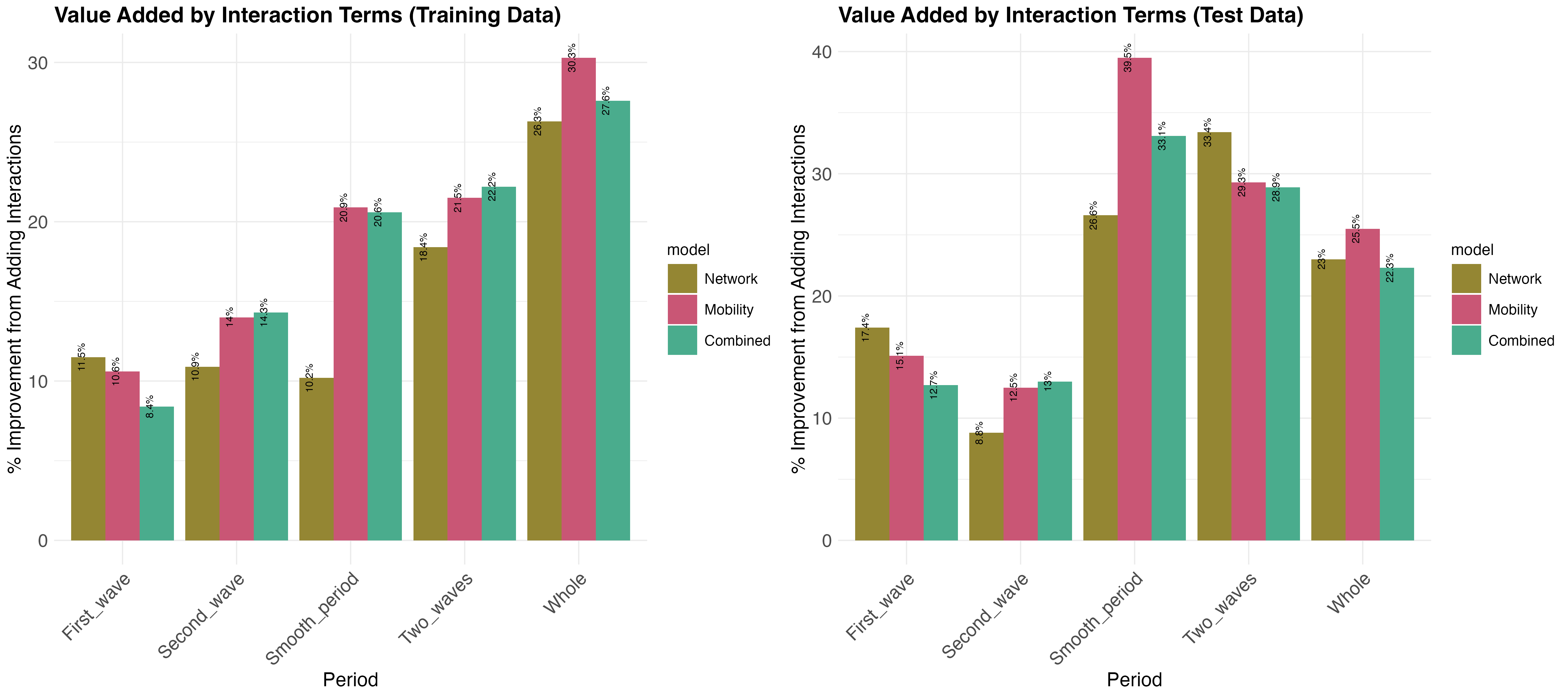}
    \caption{The added value of incorporating the interaction term in the regression model}
    \label{fig:interaction_term_value}
\end{figure*}

Figure \ref{fig:interaction_term_value} quantifies the substantial value added by interaction terms between network measures and pandemic severity. The percentage improvements range from 10-40\% across different periods, with the 'Two waves' and 'smooth' periods showing the most dramatic gains (over 30\% improvement for mobility and combined models). This highlights that the interplay between local network properties and state-level pandemic intensity is crucial, providing significant explanatory power across both periods of high transmission volatility and those with smoother, more predictable trends. The consistent positive effects across all periods and model types underscore that network measures alone are insufficient, and their interaction with pandemic context is essential for capturing transmission dynamics.

\newpage
\section{Forecasting for Longer Windows}

Table \ref{tab:forecast_horizons} shows how model performance changes with increasing prediction horizons from 1 to 4 weeks ahead. Both mobility and network measures retain significant predictive power across all forecast windows, with expected declines in coefficient magnitude as the forecast horizon extends.

\begin{table*}[!htbp]
\centering 
  \caption{Predictive Performance of COVID-19 Transmission Models at Multiple Forecast Horizons} 
  \label{tab:forecast_horizons} 
\resizebox{\textwidth}{!}{
\begin{tabular}{@{\extracolsep{5pt}}lcccccccccccc} 
\hline
                & \begin{tabular}{c} 1-Week Ahead \\ Basic Model \end{tabular} & \begin{tabular}{c} 1-Week Ahead \\ Mobility Model \end{tabular} & \begin{tabular}{c} 1-Week Ahead \\ Network Model \end{tabular} & \begin{tabular}{c} 2-Week Ahead \\ Basic Model \end{tabular} & \begin{tabular}{c} 2-Week Ahead \\ Mobility Model \end{tabular} & \begin{tabular}{c} 2-Week Ahead \\ Network Model \end{tabular} & \begin{tabular}{c} 3-Week Ahead \\ Basic Model \end{tabular} & \begin{tabular}{c} 3-Week Ahead \\ Mobility Model \end{tabular} & \begin{tabular}{c} 3-Week Ahead \\ Network Model \end{tabular} & \begin{tabular}{c} 4-Week Ahead \\ Basic Model \end{tabular} & \begin{tabular}{c} 4-Week Ahead \\ Mobility Model \end{tabular} & \begin{tabular}{c} 4-Week Ahead \\ Network Model \end{tabular}  \\
\hline
 MA & 0.003$^{***}$ & $-$0.001$^{***}$ & $-$0.0002$^{***}$ & 0.003$^{***}$ & $-$0.0005$^{***}$ & $-$0.0002$^{***}$ & 0.002$^{***}$ & $-$0.0004$^{***}$ & $-$0.0001 & 0.002$^{***}$ & $-$0.001$^{***}$ & $-$0.00002 \\ 
  & (0.0001) & (0.00004) & (0.0001) & (0.0001) & (0.00005) & (0.0001) & (0.0001) & (0.0001) & (0.0001) & (0.0001) & (0.0001) & (0.0001) \\ 
  & & & & & & & & & & & & \\ 
 intraweightt &  & $-$0.009$^{***}$ &  &  & $-$0.010$^{***}$ &  &  & $-$0.005$^{***}$ &  &  & $-$0.003$^{***}$ &  \\ 
  &  & (0.0003) &  &  & (0.0004) &  &  & (0.0005) &  &  & (0.001) &  \\ 
  & & & & & & & & & & & & \\ 
 bet &  &  & $-$5,396.817$^{***}$ &  &  & $-$6,215.035$^{***}$ &  &  & $-$8,912.698$^{***}$ &  &  & $-$8,398.193$^{***}$ \\ 
  &  &  & (577.593) &  &  & (659.776) &  &  & (711.268) &  &  & (783.620) \\ 
  & & & & & & & & & & & & \\ 
 MA:intraweightt &  & 0.00000$^{***}$ &  &  & 0.00000$^{***}$ &  &  & 0.00000$^{***}$ &  &  & 0.00000$^{***}$ &  \\ 
  &  & (0.000) &  &  & (0.000) &  &  & (0.000) &  &  & (0.000) &  \\ 
  & & & & & & & & & & & & \\ 
 MA:bet &  &  & 1.409$^{***}$ &  &  & 1.253$^{***}$ &  &  & 1.107$^{***}$ &  &  & 0.967$^{***}$ \\ 
  &  &  & (0.014) &  &  & (0.015) &  &  & (0.016) &  &  & (0.018) \\ 
  & & & & & & & & & & & & \\ 
 Constant & $-$2.950 & 41.107$^{***}$ & 17.879$^{**}$ & $-$3.251 & 48.167$^{***}$ & 18.322$^{**}$ & 0.367 & 26.794$^{***}$ & 22.409$^{**}$ & 3.946 & 19.216$^{*}$ & 23.270$^{**}$ \\ 
  & (11.661) & (6.587) & (8.277) & (11.988) & (7.917) & (9.063) & (12.239) & (9.405) & (10.156) & (12.831) & (10.760) & (11.389) \\ 
  & & & & & & & & & & & & \\ 
\hline \\[-1.8ex] 
Predict-$R^2$ & 0.5977 & 0.8909 & 0.8290 & 0.5668 & 0.7827 & 0.7203 & 0.5277 & 0.7964 & 0.7118 & 0.5391 & 0.7408 & 0.6900 \\ 
Observations & 13,271 & 13,271 & 13,271 & 13,020 & 13,020 & 13,020 & 12,770 & 12,770 & 12,770 & 12,520 & 12,520 & 12,520 \\ 
R$^{2}$ & 0.589 & 0.878 & 0.794 & 0.575 & 0.829 & 0.759 & 0.563 & 0.761 & 0.701 & 0.544 & 0.702 & 0.643 \\ 
Adjusted R$^{2}$ & 0.579 & 0.875 & 0.789 & 0.564 & 0.825 & 0.753 & 0.552 & 0.755 & 0.694 & 0.533 & 0.694 & 0.634 \\ 
\hline 
\hline \\[-1.8ex] 
\textit{Note:}  & \multicolumn{12}{l}{$^{*}$p$<$0.1; $^{**}$p$<$0.05; $^{***}$p$<$0.01} \\ 
\end{tabular}} 
\end{table*}

\begin{figure*}[!htbp]
    \centering
    \includegraphics[width=0.76\linewidth]{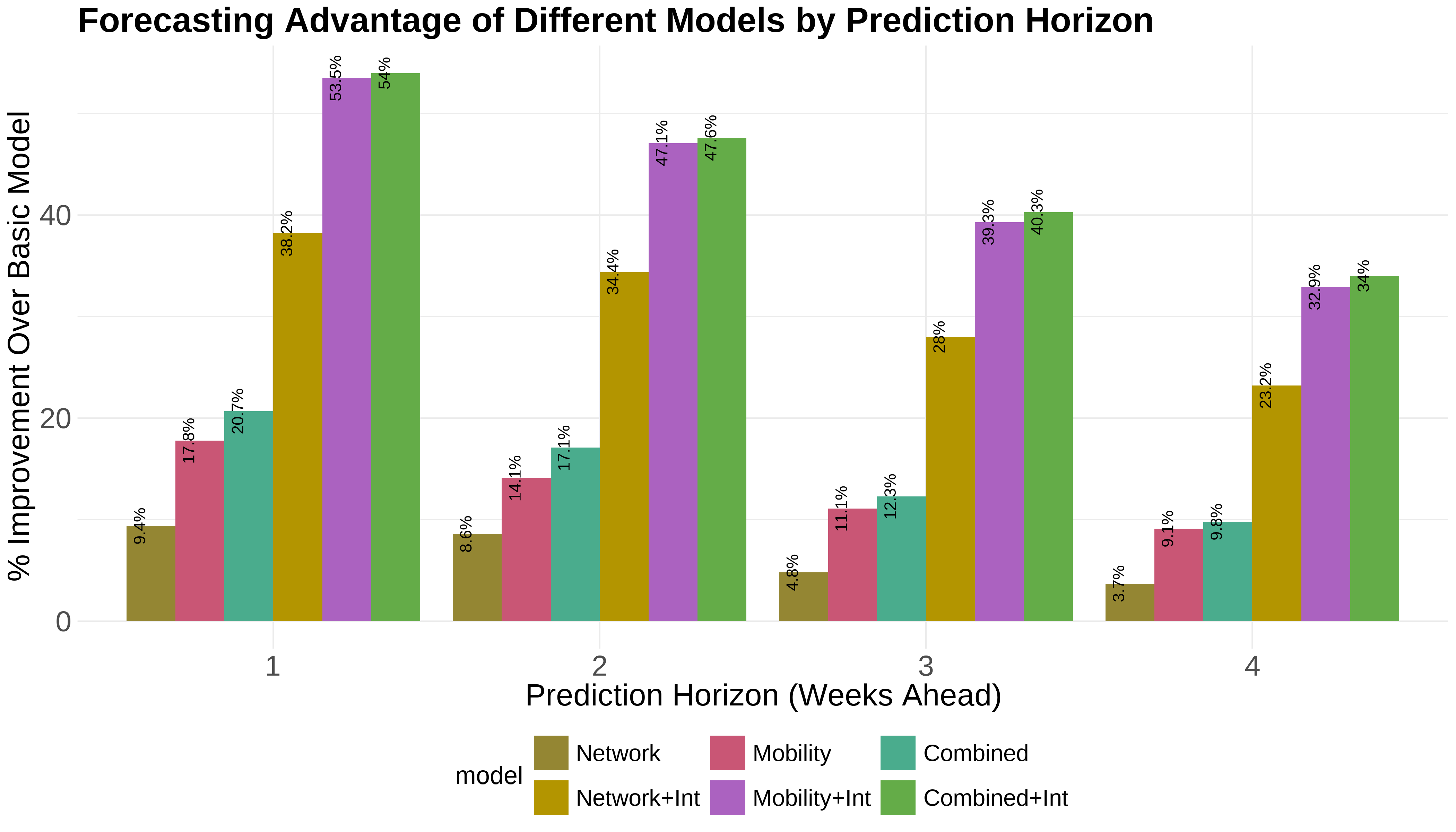}
    \caption{Prediction Horizons}
    \label{fig:forecasting_advantage_by_lag}
\end{figure*}

Figure \ref{fig:forecasting_advantage_by_lag} reveals that the forecasting advantage of network-enhanced models persists across prediction horizons, though with expected decline. At 1-week ahead forecasts, network models with interaction terms provide 35-55\% improvement over basic models, while even at 4-week horizons, these models maintain 22-32\% advantages. The consistent superiority of interaction models across all forecast windows demonstrates that network-pandemic severity interactions capture fundamental transmission mechanisms that remain predictively valuable even for longer-term forecasting. This sustained performance differential highlights the practical importance of incorporating network dynamics into pandemic analysis and response planning systems.

\section{Explainability of network interactions with and without town-level cases}

\paragraph*{Objective.}
We quantify how much variance is explained by network interactions conditional on different baselines. Let the interaction regressor be $W_{i,t}\equiv M_t Z_{i,t}$, where $M_t$ is statewide incidence and $Z_{i,t}$ is a (possibly multivariate) network exposure feature (e.g., a learned mix of \texttt{intraweightt} and \texttt{bet}). We use the Frisch--Waugh--Lovell (FWL) theorem \cite{FrischWaugh1933,Lovell1963} to derive partial-$R^2$ formulas.

\paragraph*{Setup.}
After removing town fixed effects by the within transformation, a one-week linearization gives
\begin{equation}
\widetilde{y}_{i,t+1} \;=\; \phi\,\widetilde{y}_{i,t} \;+\; \eta\,M_t\,\widetilde{Z}_{i,t} \;+\; \widetilde{\varepsilon}_{i,t+1}.
\label{eq:si_fwl_dgp}
\end{equation}
For clarity, we evaluate partial-$R^2$ in week-wise cross sections and then average over $t$.

\subsection*{Regime A: Macro-only baseline (no town histories)}
\emph{Baseline regressors}: $\{\text{FE},\, M_t\}$. \quad
\emph{Add-in regressor}: $W_{i,t}=M_t Z_{i,t}$.

Within week $t$, $M_t$ is common across towns, so the FWL residual of $W$ onto $M_t$ is
\[
W^{\perp M}_{i,t} \;=\; M_t\,\big(Z_{i,t}-\bar Z_{\cdot,t}\big),\qquad \bar Z_{\cdot,t}\equiv \tfrac{1}{N}\sum_i Z_{i,t}.
\]
The residualized outcome is
\[
y^{\perp M}_{i,t+1} \;=\; \phi\,\big(y_{i,t}-\bar y_{\cdot,t}\big)\;+\;\eta\,M_t\big(Z_{i,t}-\bar Z_{\cdot,t}\big)\;+\;\varepsilon^{\perp M}_{i,t+1}.
\]
Therefore, the partial-$R^2$ of the network interaction (averaging over weeks) is approximately
\begin{equation}
\boxed{
R^{2}_{\mathrm{part},\,\mathrm{macro}}
\;\approx\;
\frac{\eta^2\,\mathbb{E}_t\!\big[M_t^2\,\Var_i\!\big(Z_{i,t}-\bar Z_{\cdot,t}\big)\big]}
     {\eta^2\,\mathbb{E}_t\!\big[M_t^2\,\Var_i\!\big(Z_{i,t}-\bar Z_{\cdot,t}\big)\big]
      \;+\;\phi^2\,\mathbb{E}_t\!\big[\Var_i\!\big(y_{i,t}-\bar y_{\cdot,t}\big)\big]
      \;+\;\mathbb{E}_t\!\big[\Var_i\!\big(\varepsilon^{\perp M}_{i,t+1}\big)\big]}
}
\label{eq:si_partial_macro}
\end{equation}
where $\Var_i(\cdot)$ denotes cross-sectional variance across towns at week $t$.

\subsection*{Regime B: Baseline includes recent town cases}
\emph{Baseline regressors}: $\{\text{FE},\, M_t,\, y_{i,t}\}$. \quad
\emph{Add-in regressor}: $W_{i,t}=M_t Z_{i,t}$.

Within week $t$, let $Z^{\,\perp y}_{i,t}=Z_{i,t}-\bar Z_{\cdot,t}-\gamma_t(y_{i,t}-\bar y_{\cdot,t})$ be the cross-sectional residual of $Z$ on $y$ (with slope $\gamma_t$). Then
\[
W^{\perp (M,y)}_{i,t} \;=\; M_t\,Z^{\,\perp y}_{i,t},\qquad
y^{\perp (M,y)}_{i,t+1} \;=\; \eta\,M_t\,Z^{\,\perp y}_{i,t} \;+\; \varepsilon^{\perp (M,y)}_{i,t+1}.
\]
Therefore, the partial-$R^2$ reduces to
\begin{equation}
\boxed{
R^{2}_{\mathrm{part},\,\mathrm{with\,town}}
\;\approx\;
\frac{\eta^2\,\mathbb{E}_t\!\big[M_t^2\,\Var_i\!\big(Z^{\,\perp y}_{i,t}\big)\big]}
     {\eta^2\,\mathbb{E}_t\!\big[M_t^2\,\Var_i\!\big(Z^{\,\perp y}_{i,t}\big)\big]
      \;+\;\mathbb{E}_t\!\big[\Var_i\!\big(\varepsilon^{\perp (M,y)}_{i,t+1}\big)\big]}
}
\label{eq:si_partial_with}
\end{equation}
with $\Var_i(Z^{\,\perp y}_{i,t})=\Var_i(Z_{i,t}-\bar Z_{\cdot,t})\,[1-R^2_{Z\mid y}(t)]$.

\paragraph*{Implications.}
\begin{itemize}
\item \emph{Prevalence activation}: Both \eqref{eq:si_partial_macro} and \eqref{eq:si_partial_with} scale with $M_t^2$, so explainability is largest in high-incidence weeks (waves/rising phases).
\item \emph{Orthogonalization loss}: With town histories, the usable cross-sectional variance of the network exposure is reduced by $(1-R^2_{Z\mid y}(t))$, shrinking the numerator in \eqref{eq:si_partial_with}.
\item \emph{Headroom loss}: With strong carryover $\phi$, the baseline with $y_{i,t}$ explains most one-week variation; remaining variance for networks to explain is small.
\end{itemize}

\paragraph*{Extensions.}
If $Z_{i,t}=X_{i,t}^\top \theta$ (a combination of several centralities), replace $\Var_i(Z)$ by the variance of the combined exposure and $\Var_i(Z^{\,\perp y})$ by its residual variance after orthogonalizing on $y_{i,t}$ (within week).

\bibliographystyle{unsrt}
\bibliography{mobility_refs}